\documentclass[prb,aps,twocolumn,superscriptaddress]{revtex4}
\usepackage{lscape, graphicx}
\usepackage{amsmath,amssymb,amsfonts,bm}
\usepackage{longtable}
\usepackage{subfigure}
\usepackage{lscape, graphicx}
\usepackage{color}
\usepackage{comment}

\newcommand{\be}{\begin{equation}}
\newcommand{\ee}{\end{equation} }
\newcommand{\ba}{\begin{eqnarray}}
\newcommand{\ea}{\end{eqnarray} }
\newcommand{\n}{\nonumber \\ }

\begin{document}
\input{epsf.sty}

\title{Leggett mode in a strong-coupling model of iron arsenide superconductors}

\author{F. J. Burnell}
 \affiliation{Department of Physics,  Princeton University, Princeton, NJ 08544}
 \affiliation{Kavli Institute for Theoretical Physics, Santa Barbara, CA}
 \affiliation{All Souls College, Oxford, OX14AL}

\author{Jiangping Hu}
\affiliation{Department of Physics, Purdue University, West
 Lafayette, Indiana 47907}

\author{Meera M. Parish}
\affiliation{Princeton Center for Theoretical Science, Princeton
  University, Princeton, NJ 08544}
\affiliation{Cavendish Laboratory, JJ Thomson Avenue, Cambridge, CB3
0HE, UK}

\author{B. Andrei Bernevig}
 \affiliation{Department of Physics,  Princeton University, Princeton, NJ 08544}
\affiliation{Princeton Center for Theoretical Science, Princeton
  University, Princeton, NJ 08544}

\begin{abstract}
Using a two-orbital model of the superconducting phase of the
pnictides, we compute the spectrum of the Leggett mode -- a
collective excitation of the phase of the superconducting gap known
to exist in multi-gap superconductors -- for different possible
symmetries of the superconducting order parameter.  Specifically, we
identify the small regions of parameter space where the Leggett mode
lies below the two-particle continuum, and hence should be visible
as a sharp resonance peak.  We discuss the possible utility of the
Leggett mode in distinguishing different momentum dependencies of
the superconducting gap.  %We show that the existence of an undamped
%Leggett mode requires large band renormalization factors. Thus, the
We argue that the observation of a sharp Leggett mode would be
consistent with the presence of strong electron-electron
correlations in iron-based superconductors. We also emphasize the
importance of the orbital character of the Leggett mode, which can
result in an experimental observation of the mode in channels other
than $A_{1g}$.
\end{abstract}

\date{\today}

%\pacs{72.25.-b, 72.10.-d, 72.15. Gd}

\maketitle

\section{Introduction} \label{Introduction}
The discovery of high-temperature superconductivity in iron arsenide
and related compounds at the beginning of 2008~\cite{kamihara2008}
has triggered an enormous interest in the condensed matter physics
community and has stimulated a flurry of experimental
activity~\cite{kamihara2008,takahashi2008,ren2008,GFchen2008,XHchen2008,wen2008,rotter2008,sasmal2008,GFchen2008_3}.
%The iron-based superconductors form a wide family of materials, and
%while most of them exhibit both a magnetic and a superconducting
%phase, there are several notable \textbf{(cite)} members of this
%family that only exhibit one of these phases without the other.
Upon electron~\cite{Chen2008} or hole~\cite{rotter2008} doping of a
magnetically-ordered parent state, most of the iron-based
superconductors exhibit transition temperatures $T_c$ beyond the
conventional BCS regime, with some extending up to 56
K~\cite{chen-08n761}, thereby breaking the cuprate monopoly on
high-temperature superconductivity. Experimental evidence
accompanied by theoretical modeling suggest that the pairing in the
iron-pnictides is different from the $d$-wave pairing of the
cuprates. Nevertheless, they resemble the cuprates in that it is
increasingly clear that the magnetism of the parent state (either
long-range or fluctuating order) crucially influences the pairing
symmetry of the doped system. A conclusive observation of the
pairing symmetry still remains elusive, with both nodal and nodeless
order parameters reported in experiments.  This provides a strong
incentive to identify new experimental probes potentially sensitive
to the symmetry of the superconducting gap.

While a wide range of nodal gap functions were initially
predicted~\cite{si2008,lee2008,raghu2008}, the general theoretical
view has now converged to favor an extended $s$-wave order parameter
(denoted $s^\pm $ or $s_{x^2 y^2}$) that takes opposite signs on the
electron and hole pockets along the multi-band Fermi surfaces.  The
symmetry of this $s_{x^2 y^2}$ gap matches that  of the
iron-pnictide Fermi surface: it is maximal around $(0,0), (\pi,0),
(0,\pi), (\pi,\pi)$ - the location of the Fermi surfaces in the
unfolded one-iron-per-site Brilloiun zone.  This sign-alternating
nodeless gap is consistent with some experimental data and also has
broad theoretical
support~\cite{mazin2008,kuroki2008,seo2008,wang2009,Parish2008,thomalefrg2009,Parker2008b,chubukov2008A,Cvetkovic2009}.
Indeed,  both strong\cite{seo2008,Parish2008}- and
weak\cite{mazin2008,kuroki2008,wang2009,thomalefrg2009,Parker2008b,chubukov2008A,Cvetkovic2009, thomale2010, stanev2008,Daghofer2010}-
coupling  theories of the onset of superconductivity predict an
extended $s$-wave order parameter.

Experimentally, however, there is no consensus about the nature of
the order parameter, with both nodal and nodeless gaps being
reported. While most experiments can be explained within the
framework of an $s^{\pm}$ gap~\cite{Mazin2009}, several facts, such
as the $T^3$ dependence of the NMR relaxation rate over a
significant temperature range~\cite{grafe2008,nakai2008,Mukuda2008},
residual finite quasiparticle terms in the thermal
conductivity~\cite{Tanatar2008,Check2008},  as well as the power-law
behavior of the penetration depth~\cite{martin2008,hashimoto2008},
remain unsettled.
Some of the experiments on penetration depth and thermal conductivity
could be explained by an $s^{\pm}$ order parameter if there were a
large gap anisotropy \cite{martin2008,hashimoto2008}, but this
contradicts ARPES data, which reveals very
isotropic nodeless gaps on the hole Fermi
surfaces~\cite{Hasan2008,Zhou2009a,Feng2008a}, of
magnitudes matching a strong-coupling form $\Delta(k) = \Delta_0
\cos(k_x) \cdot \cos(k_y)$~\cite{seo2008} in the unfolded Brillouin
zone.

A possible resolution of this apparent contradiction, consistent
with the theoretical prediction of an $s_\pm$ order parameter, is
that the gap anisotropy is doping dependent and that different
experiments are done at different dopings. In the strong-coupling
mean-field picture~\cite{seo2008,Parish2008}, the gap anisotropy is
intrinsically doping dependent: the gap has a form $\cos(k_x) \cdot
\cos(k_y)$ which becomes more anisotropic as the doping is
increased.  In a weak-coupling expansion of Fermi surface %(FS)
interactions, the gap anisotropy can arise from the presence of an
$A_{1g}$ term $\cos(k_x) + \cos(k_y)$ (which does not break the
crystal symmetry but can create nodes on the $(\pi,0)$ and $(0,\pi)$
electron surfaces) in the band interactions \cite{vorontsov2009}
upon renormalization \cite{thomalefrg2009}.  A large gap anisotropy
is already present in Functional Renormalization Group studies of
orbital models~\cite{wang2009, thomale2010}.

In this paper, we analyze another physical phenomenon -- the Leggett
mode of multi-band superconductors --  that depends on the strength
of the pairing order parameter and could also in principle
quantitatively distinguish a sign-changing gap from other gap
symmetries. Specifically, we investigate to what extent the pairing
symmetry of the iron-based superconductors can be deduced by
analyzing the behavior of the Leggett mode as a function of doping
and the strength of superconducting order parameters.  As the
iron-based superconductors have multiple orbitals, the
superconducting state exhibits a plethora of collective modes beyond
the usual Goldstone/Higgs plasmon.  Here we use an effective
two-orbital model of the superconducting state to study one of these
-- the Leggett mode associated with anti-symmetric phase
fluctuations between the two superconducting order parameters. This
gapped collective mode can, in the right parameter range, present a
sharp collective mode resonance below the two-particle continuum
which could in principle be detected experimentally.

To determine whether such a collective mode resonance occurs in the
pnictide superconductors, we study the gap and dispersion of the
Leggett mode as a function of doping and the superconducting order
parameters. We show that, for a sign-changing gap function, the
Leggett mode can be below the two-particle continuum for a small
regime at low doping. In particular, when the band renormalization
is large, an undamped Leggett mode can exist in a relatively large
parameter region. Thus, the observation of a sharp Leggett mode will
validate the presence of strong electron-electron correlations in
the iron-based superconductors. Moreover, in our two-orbital model,
the Leggett mode is a $B_{1g}$ mode, instead of a pure $A_{1g}$
mode, which is expected in any band-based model. Therefore, the
orbital structure of pairing in the iron-based superconductors can
be validated by identifying the existence of the Leggett mode in
channels other than $A_{1g}$.

Unfortunately, we find that the Leggett mode cannot qualitatively
distinguish between a sign-changing order parameter and other gapped
order parameters. However, 
the sign-changing order parameter will have a degree of anisotropy
which depends on doping. For large doping, the sign-changing order
parameter on the enlarged Fermi surface will exhibit larger
anisotropy. As such, the superconducting gap will be small on some
parts of the Fermi surface and the Leggett mode will be overdamped,
lying above the two-particle continuum, and hence unobservable. This
presents a testable opportunity if, at moderate doping (when the
gaps should theoretically be isotropic), the Leggett mode is below
the two-particle continuum and hence observable. If so, the
observation of a disappearing collective mode provides indirect
support for a sign-changing gap function.

\section{Computing the effective action} \label{CalcSect}
The Leggett mode is a collective excitation of two- (or multi-) band
superconductors, associated with anti-symmetric phase oscillations
between the two bands.  It is thus a neutral mode associated with
oscillations between the supercurrents of the two bands.  Here we
present the effective action for this mode derived from  a two-orbital
model appropriate to the pnictides at temperatures well below the
onset of superconductivity.
To render our
calculations analytically tractable, we focus on a simplified model
of the iron-based superconductors that takes into account only the
$d_{xz}, d_{yz}$ orbitals. In this case, an intra-band order
parameter can have its phase fluctuate between the two orbitals in
two modes: the usual symmetric combination (Goldstone) and the
antisymmetric combination, which is the Leggett mode.

While the conventional Leggett mode involves only the Fermi surface
gaps, our work involves a Leggett mode in the orbital gaps.
We consider the orbital basis rather than an effective band basis,
because neglecting the orbital structure of the iron-based
superconductors is most likely incorrect: it was shown
\cite{ran2009} that due to the difference in mirror symmetry
eigenvalues
of the electron and one of the  hole bands at the $\Gamma$ point in
the Brillouin zone (BZ), the spin density wave (SDW) state is
gapless with a Dirac point in both two and five-orbital models of
iron-based superconductors. This highly nontrivial effect, confirmed
by experiments \cite{hsieh2008}, is lost in the effective band-basis
picture.  Details of the derivation of the Leggett mode effective
action in the orbital basis, which differs slightly from the
band-basis result of, e.g., Ref.~\onlinecite{Iskin}, are given in
Appendix~\ref{DerivationApp}.

\subsection{Model Hamiltonian}

Using the insight provided by numerical and analytic studies
suggesting that the antiferromagnetic exchange coupling between
next-nearest-neighbor Fe sites is strong~\cite{yildirim2008,si2008},
two of us \cite{seo2008} studied  a $t$-$J_1$-$J_2$ model without
band renormalization and obtained a gap function of the form
$\cos(k_x) \cdot \cos(k_y)$, which changes sign between the electron
and hole-pockets of the Fermi surface of the material.
It is this type of strong-coupling superconductivity that we will
focus on  in this paper, but we point out that other weak coupling
approaches exist and give a similar sign-changing order parameter
\cite{mazin2008,kuroki2008,wang2009,thomalefrg2009,Parker2008b,chubukov2008A,Cvetkovic2009}.

To calculate the effective action for the phase modes of the
superconducting state, we employ a model of the pnictides which
incorporates only the  
$d_{xz}$ and $d_{yz}$ orbitals at each site, together with
hybridization between the two. Although this description is only
truly valid in the case of unphysically large crystal field
splitting, we use this model for its analytic simplicity. We adopt
the band structure proposed in Ref.~\onlinecite{raghu2008}, which at
first glance captures the essence of the Density Functional Theory
results:
 \begin{eqnarray} \label{Eq_KinHam}
&H_0 =
\sum_{k\sigma}\psi^\dagger_{k\sigma} T(k) \psi_{k\sigma} + H_{int} \nonumber \\
 &T(k)=\left(\begin{array}{cc}
 \epsilon_x(k) & \epsilon_{xy}(k)\\
 \epsilon_{xy}(k) & \epsilon_y(k)
 \end{array}\right)
\end{eqnarray}
where $\psi_{k,\sigma}^\dagger = (c^\dagger_{d_{xz},k,\sigma},
c^\dagger_{d_{yz},k,\sigma}) $ is the creation operator for spin
$\sigma$ electrons in the two orbitals and the kinetic terms read:
\ba
 &\epsilon_x(k) =  -2t_1\cos k_x - 2t_2\cos k_y -4t_3 \cos k_x \cos k_y  -\mu \nonumber \\
 &\epsilon_y(k) = -2t_2\cos k_x - 2t_1\cos k_y -4t_3 \cos k_x \cos k_y  - \mu \nonumber  \\
 &\epsilon_{xy}(k) = -4t_4\sin k_x \sin k_y \label{dispersion1}
\ea
The hoppings have roughly the same magnitude: $t_1 = -1.0, t_2=1.3,
t_3=-0.85,$ and $t_4 = -0.85$ in eV.  We find that the half-filled, two
electrons per site configuration is achieved when $\mu=1.54 eV$.

The missing ingredient in this two-orbital model is the $d_{xy}$
orbital, which can be shown to be important to the detailed physics
of the iron-based superconductors \cite{graser2008}. For example,
the kinetic model (\ref{Eq_KinHam}) gets the location of the second
hole pocket wrong -- it situates it at the $(\pi,\pi)$ point in the
unfolded BZ, whereas LDA calculations show two hole pockets at the
$\Gamma$ point.  However, the two-orbital model gets several of the
qualitative characteristics of the iron-based superconductors right:
it has a nodal SDW instability and a sign-changing $s$-wave
superconducting instability.

To describe the superconducting phase, we use the approach of
Ref.~\onlinecite{seo2008}, adopting a strong-coupling picture in
which the interaction Hamiltonian %$H_{int}$
contains anti-ferromagnetic nearest-neighbor and
next-nearest-neighbor coupling between the spins in both identical
and opposite orbitals. While not entirely correct at lattice scales,
it was shown that this model gives remarkably large overlaps with
the interactions obtained through the functional renormalization
group \cite{wang2009A} method, and hence can be considered as an
effective interaction model for the iron-based superconductors.
Furthermore, for our purposes these interactions are important only
insofar as they give, after decoupling in the superconducting
channel, the sign-changing $\cos(k_x) \cdot \cos(k_y)$
superconducting order parameter. In this sense, the interacting
spin-model we use can be thought of as an effective Ginzburg-Landau
description of iron-based superconductors; the precise mechanism
driving the transition to the superconducting phase is irrelevant to
the effective action we derive here.

Based on the mean-field analysis of Ref.~\onlinecite{seo2008}, we
will assume throughout that the superconducting instability is
dominated by the intra-orbital interactions, so that the gap is
diagonal in the orbital basis. Indeed, at the mean-field level, the
inter-orbital pairing is weaker than the intra-orbital pairing by a
factor of approximately five\cite{seo2008}.  In addition there is a
large on-site inter-orbital Hund's rule coupling which will not
enter into the present analysis as it does not alter the nature of
the order parameter at mean-field level. We will briefly discuss the
impacts of this last term, together with the antiferromagnetic
inter-orbital interactions, in Sect.~\ref{Sect_OtherInts}.

\subsection{Phase-only effective action}

To obtain an effective action for the phase of the superconducting
gap, we follow the general protocol of
Ref.~\onlinecite{SharapovBeck2}; details of this calculation as
applied to the orbital basis are given in
Appendix~\ref{DerivationApp}.  In essence, one first decouples the
interaction terms in the microscopic model using a
Hubbard-Stratonovich transformation.  This re-expresses operators
quadratic in the fermions as interaction terms between a pair of
fermions and the superconducting field $\Phi$. Deep in the
superconducting region, where fluctuations in the magnitude $\Delta
\equiv | \Phi|$ can be neglected, integrating out the fermions then
yields an effective action for the phase modes of the system. Since
we work with a two-orbital model, there are a priori two
superconducting gaps, excluding inter-orbital pairing.
Though by symmetry their magnitudes have to be equal, this leads to
two independent phase degrees of freedom.  As is well known, one of
these is a Goldstone mode which, upon including the Coulomb
interactions, becomes a plasma mode.  The other is the (gapped)
Leggett mode, which will be our principle focus here.

For our purposes, the two phase degrees of freedom are most
conveniently expressed in the basis
\ba
\phi \equiv \frac{1}{\sqrt{2}} \left( \theta_1 + \theta_2 \right ) & \ \ \ \ \ \
\varphi \equiv \frac{1}{\sqrt{2}} \left( \theta_1 - \theta_2 \right )
\ea
where $\theta_1$ and $\theta_2$ are the phases of the gaps in the
$xz$ and $yz$ orbitals, respectively.   Hence $\phi$ represents the
symmetric phase oscillation, while $\varphi$ represents the
(neutral) antisymmetric phase mode. In this basis, we find the
effective action to be (see calculation details in
Appendix~\ref{DerivationApp}):
\begin{widetext}
\ba \label{Eq_SNoC} S_{eff} &=& \int d \Omega d^2  q \left (
\begin{array}{cc} \phi(\Omega, q) &  \varphi(\Omega, q) \end{array} \right) \left (
\begin{array}{cc} N_{\phi \phi}\left[ \Omega^2- c_{\phi\phi, ij}^2
q_i q_j \right ] &
c_{\phi \varphi, ij}^2 q_iq_j \\
c_{\phi \varphi , ij}^2 q_iq_j & N_{\varphi \varphi}\left[ \Omega^2
  - \Omega_0^2 - c_{\varphi \varphi, ij}^2 q_i q_j\right]  \\
\end{array} \right)
\left ( \begin{array}{cc} \phi(\Omega, q) \\  \varphi(\Omega, q) \\
 \end{array} \right)   \ \ \ \ .
\ea
with momentum-independent coefficients given by:
 \ba \label{Eq_Ntot}
  N_{\phi \phi} &=& - \int \frac{d^2 k}{ (2
\pi)^2} \left \{ \frac{\Delta^2}{4 E^{(\Delta) 3}_{+} } +
\frac{\Delta^2}{4 E^{(\Delta) 3}_{-} } \right \}\\
N_{\varphi \varphi} &=&
  -\int \frac{d^2 k}{(2\pi)^2}  \frac{(\epsilon_{x}-\epsilon_{y})^2}
  {(E_+-E_-)^2 }\left \{ \frac{ \Delta^2}{ 4E^{(\Delta)  3}_{+}}
  + \frac{ \Delta^2}{4 E^{(\Delta)  3}_{-}} \right \} \n
& & - \int \frac{d^2 k}{(2\pi)^2} \frac{8 \Delta^2
  \epsilon_{xy}^2}{(E_+-E_-)^2}
\frac{E_+^2+E_-^2+\Delta^2+E_+^{(\Delta)}E_-^{(\Delta)}-E_+ E_-
}{(E^{(\Delta)}_+ +E^{(\Delta)}_-)^3 E^{(\Delta)}_{+}
  E^{(\Delta)}_{-}}  \\
M &=&  \int \frac{d^2 k}{ (2 \pi)^2} \frac{4 \Delta^2
    \epsilon_{xy}^2}{ E_+^{(\Delta)}E_-^{(\Delta)}
    (E_+^{(\Delta)}+E_-^{(\Delta)})}  \label{Eq_Mtot0}
 \ea
with $\Omega_0 \equiv \sqrt{\frac{M}{- N_{\varphi \varphi}}}$. Here,
$E_{\pm}$ are the two band energies $ E_{\pm} = \frac{1}{2} \left(
\epsilon_{x} + \epsilon_{y} \pm \sqrt{ \left(\epsilon_{x}
  - \epsilon_{y} \right)^2 + 4 \epsilon_{xy}^2} \right ) $ of the metallic state, and
$E_{\pm}^{(\Delta)}= \sqrt{ E_{\pm}^2+ \Delta^2}$ are the
quasi-particle energies in the superconducting phase.  All
$\epsilon, E,$ and $\Delta$ are evaluated  at the momentum $k$ to be
integrated over.  The above equations represent the main result of the paper.

In Eq.~(\ref{Eq_SNoC}), terms linear in $q$, as well as terms
bilinear in $q, \Omega$, all vanish in the limit $T \rightarrow 0$.
As expected, this effective action (\ref{Eq_SNoC}) describes one
gapless mode, comprised entirely of symmetric phase fluctuations at
$q=0$, and one gapped mode.  The latter is the Leggett mode; at
$q=0$ it consists purely of antisymmetric phase oscillations between
the two superconducting gaps.  Here we are principally interested in
the Leggett mode gap, $\Omega_0$, as this represents the threshold
at which the mode becomes experimentally observable. Thus if
$\Omega_0< 2 \Delta$, we expect the Leggett mode to appear as a
sharp resonance in the spectrum of the pnictide superconductors.

For terms involving $q^2$, the expressions for the coefficients
$c_{\alpha\beta, ij}^2$ are somewhat more complicated and are thus
given in Appendix \ref{TraceApp}. We note, however, that  for $i
\neq j$, any coefficient of $q_i q_j$ vanishes due to symmetry.
Further, for $i=j$, symmetry of the coefficients under a $90$ degree
rotation of the Brillouin zone fully determines their direction
dependence in $q$.
Taking these symmetries into account,  Eq. (\ref{Eq_SNoC}) has the
form:
\ba %\label{Eq_SNoC}
S_{eff} &=& \int d \Omega d^2  q \left ( \begin{array}{cc} \phi &  \varphi \end{array} \right)
\left ( \begin{array}{cc}
N_{\phi \phi}\left[ \Omega^2- c_{\phi\phi}^2 q^2 \right ] & c_{\phi \varphi}^2 (q_x^2-q_y^2) \\
c_{\phi \varphi}^2 (q_x^2-q_y^2) & N_{\varphi \varphi}\left[ \Omega^2
  - \Omega_0^2 - c_{\varphi \varphi}^2 q^2\right] 0 \\
\end{array} \right)
\left ( \begin{array}{cc} \phi \\  \varphi \\
 \end{array} \right)   \ \ \ \ .
\ea
\end{widetext}

We should note that the Leggett mode gap is proportional to
$\epsilon_{xy}^2$ -- that is, to the off-diagonal kinetic terms in
the orbital basis.  This is in contrast to the approach of, for
example, Ref.~\onlinecite{Iskin}, in which the superconducting gap
is taken to be diagonal in the band basis of the normal state, and
it is the inter-band interactions which couple the phases of the two
gaps, and hence generate the Leggett mode. This difference stems
from the fact that we take the gap to be diagonal in the orbital
basis: $\Delta_{\alpha} (k) = \langle c_{\alpha \uparrow k}
c_{\alpha \downarrow -k} \rangle $, where $\alpha$ indexes the
orbitals and assume that the pairing is defined over the whole
Brillouin Zone.  Any  model in which the interaction is written in
orbital space  and which aims to respect the point-group symmetries
of the  lattice will require this type of orbital-basis formalism.

\subsection{Including Hunds interactions}  \label{Sect_OtherInts}

In light of the fact that our approach is based on an absence of
off-diagonal interactions in the superconducting channel (when the orbital basis is used), it
is useful to consider in more detail the validity of this assumption
in the presence of inter-orbital couplings. In the pnictides the
ferromagnetic Hund's rule interaction
 \be
 H_H = - J_H \sum_r S_{1 r}
S_{2 r} \equiv -J_H \sum_r c^\dag_{1 r \sigma} \sigma_{\sigma, \sigma'}
c_{1 r \sigma'}  c^\dag_{2 r \gamma} \sigma_{\gamma, \gamma'} c_{2 r
\gamma'}
 \ee
is the principal source of such interactions.

Since spin ordering must be absent in the superconducting 
phase, generically we may decouple the Hunds interaction in either the 
particle-particle channel or the particle-hole channel.  At lowest loop order, 
the particle-particle interaction serves only to renormalize the band structure.  
The particle-hole contribution was, as previously noted, shown to be small by Ref. 
\onlinecite{seo2008}.  Neglecting the small inter-orbital pairing at mean-field, 
we find that the Hunds interaction affects the effective action for the Leggett mode
$\varphi$ only through higher loop corrections in the fermion
propagator.

Further, it is straightforward to include the effect of the small inter-orbital 
interaction  in the
superconducting channel.  Such a term simply modifies the effective action for 
the superconducting phase by adding a term $ V_{12} \left( \Delta_1\Delta_2^* + \Delta_2 \Delta_1^* \right)
\equiv 2 |\Delta |^2 V_{12}\cos( \varphi)$.  This modifies the gap of the Leggett mode according to
\be \label{Eq_LegBAnd}
\Omega_0^2 \rightarrow \Omega_0^2  -  \frac{V_{12}}{ V_{11}V_{22}
- V_{12}^2} \frac{\Delta_0^{2} }{N_{\varphi \varphi}} \ \ \ .
 \ee
Here $V_{\alpha \beta}$ parametrize the superconducting interaction between orbitals $\alpha$ and $\beta$, 
as described in Eq. \ref{Eq_FProp}, and we have taken $\Delta(k) = \Delta_0 \cos(k_x) \cos(k_y)$.  For $0< V_{12} \ll V_{11}, V_{22}$, the effect of including such a term is always to bring the Leggett mode gap down in energy.

\subsection{Effective action with Coulomb terms}
\label{Sect_Coulomb}

In the above analysis, we ignored the effects of the Coulomb
interaction on the phase modes.  In a single band superconductor,
including the Coulomb interactions modifies the effective action for
the phase $\theta$ of the superconducting gap such that $\theta$
becomes a plasma mode \cite{PhysRev.112.1900}. We will not examine
in detail the plasma mode here; rather we note that including
Coulomb interactions does not substantially modify the relevant
features of the Leggett mode, as we show below.  This is not
surprising since the Leggett mode is, at long wavelengths,
associated with the neutral antisymmetric phase oscillations,
 and hence
does not couple to the Coulomb interaction.

In the presence of Coulomb interactions, the phase-only effective
action has the form:
\begin{widetext}
\ba \label{SeffCoul}
 S_{eff} &=& \int d \Omega d^2 k \left (
\begin{array}{cc} \phi & \varphi\end{array} \right) \left (
\begin{array}{cc} N_{\phi \phi}\left[ \frac{\Omega^2}{1- U(q)
N_{\phi \phi} }- c_{\phi \phi}^2 q^2
\right ] & c_{\phi \varphi}^2 ( q_x^2-q_y^2)\\
c_{\phi \varphi}^2( q_x^2-q_y^2) & N_{\varphi \varphi}\left[ \Omega^2
  - \Omega_0^2 - c_{\varphi \varphi}^2  q^2\right] \\
\end{array} \right)
\left ( \begin{array}{cc} \phi \\  \varphi \\
 \end{array} \right)  \n
&&+ \Delta_\alpha V^{-1} _{\alpha, \beta} \Delta_\beta
 \ea
(For the sake of completeness we
derive this result in Appendix \ref{CoulombApp}).
The dispersion of the symmetric mode $\phi$ is modified by the
denominator of its $\Omega^2$ term; in practice since the Coulomb
interaction $U(q)$ is singular as $q\rightarrow 0$ this makes the
symmetric mode into a plasma mode, exactly as is known to occur in a
single band superconductor~\cite{SharapovBeck}.

Eq.~\eqref{SeffCoul} shows that including Coulomb interactions does
not alter the mass gap of the Leggett mode, as the plasma mode does
not mix with the Leggett mode $\varphi$ at $q=0$.  The net effect of
the Coulomb terms on $\varphi(\Omega, q)$ will be a modification of
the $q^2$ term in the effective action of the Leggett mode.
Integrating out $\phi$, we obtain:
\be \label{Eq_SEffFinal}
S_{eff} = \int d \Omega d^2 k
\varphi(q) \varphi (-q) N_{\varphi \varphi}\left[ \Omega^2 -
\Omega_0^2 - c_{\varphi \varphi}^2 ( q_x^2-q_y^2) \left( 1 +
  \frac{c_{\phi \varphi}^4}{c_{\varphi \varphi}^2 N_{\varphi \varphi} }
  \frac{q_x^2-q_y^2}{\frac{\Omega^2}{( 1- N_{\phi \phi} U(q) )} + q^2 c_{\phi \phi}^2 }
\right ) \right] + \Delta_\alpha V^{-1} _{\alpha, \beta}
\Delta_\beta
 \ee
Hence for small $q, \Omega$, in the presence of Coulomb
interactions, provided that $\lim_{q \rightarrow 0}
\frac{\Omega^2}{q^2( 1- N_{\phi \phi} U(q) )}$ is finite, the net
effect is a modification of the effective velocity of the mode. The
above equation needs to be solved self-consistently to obtain the
mode dispersion. However, the limit $q\rightarrow 0$, which
determines whether the Leggett mode is above or below the
particle-particle continuum, is unchanged from the case without the
Coulomb interaction.
\end{widetext}

%\begin{widetext}
\begin{figure*}[htp]
\centering
% \begin{center}
 \subfigure[]{
    \includegraphics[width=2.75in]{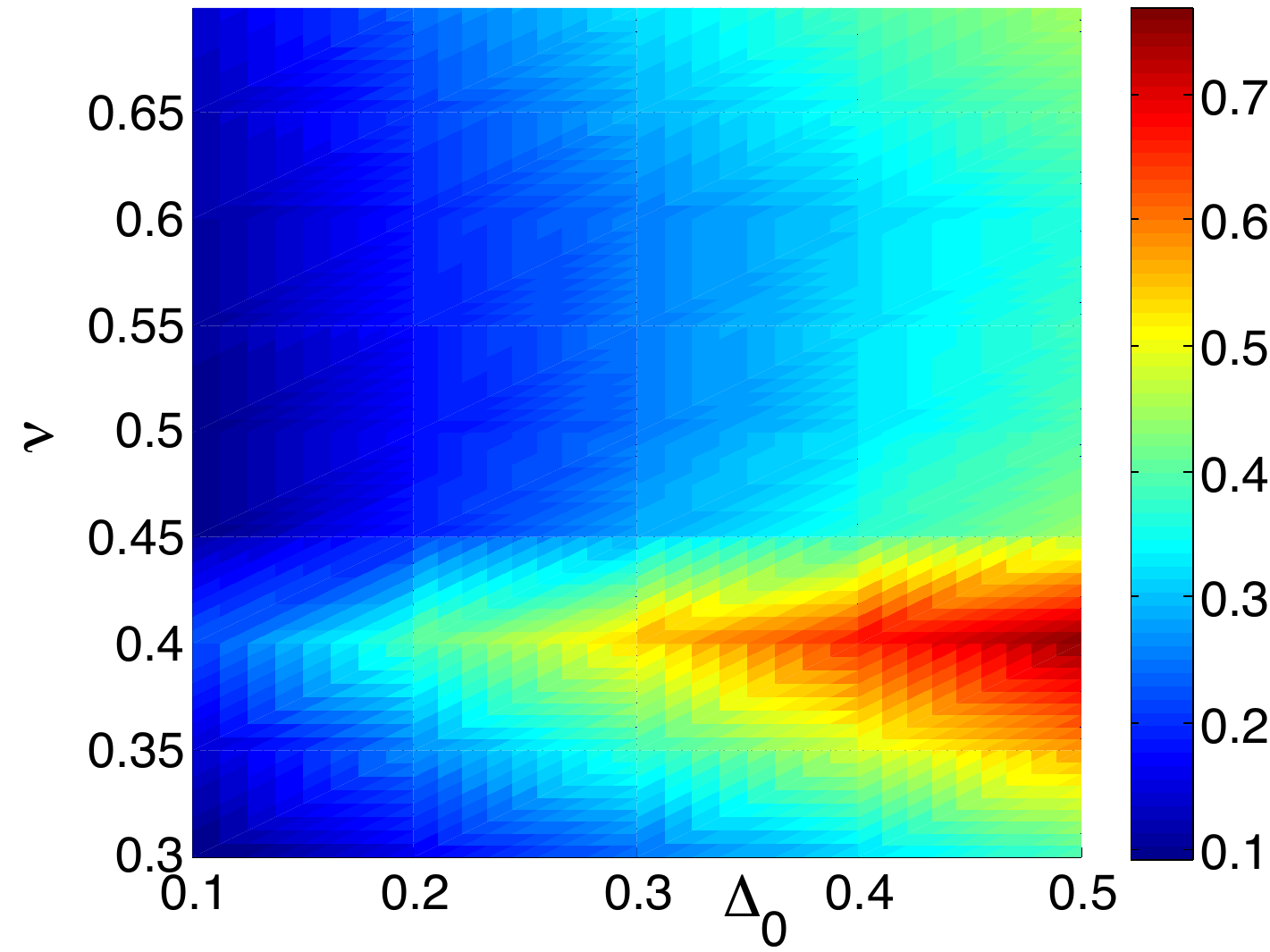}
    \label{ms1}
} \hspace{1cm}
\subfigure[] % caption for subfigure a
{
    \includegraphics[width=2.75in]{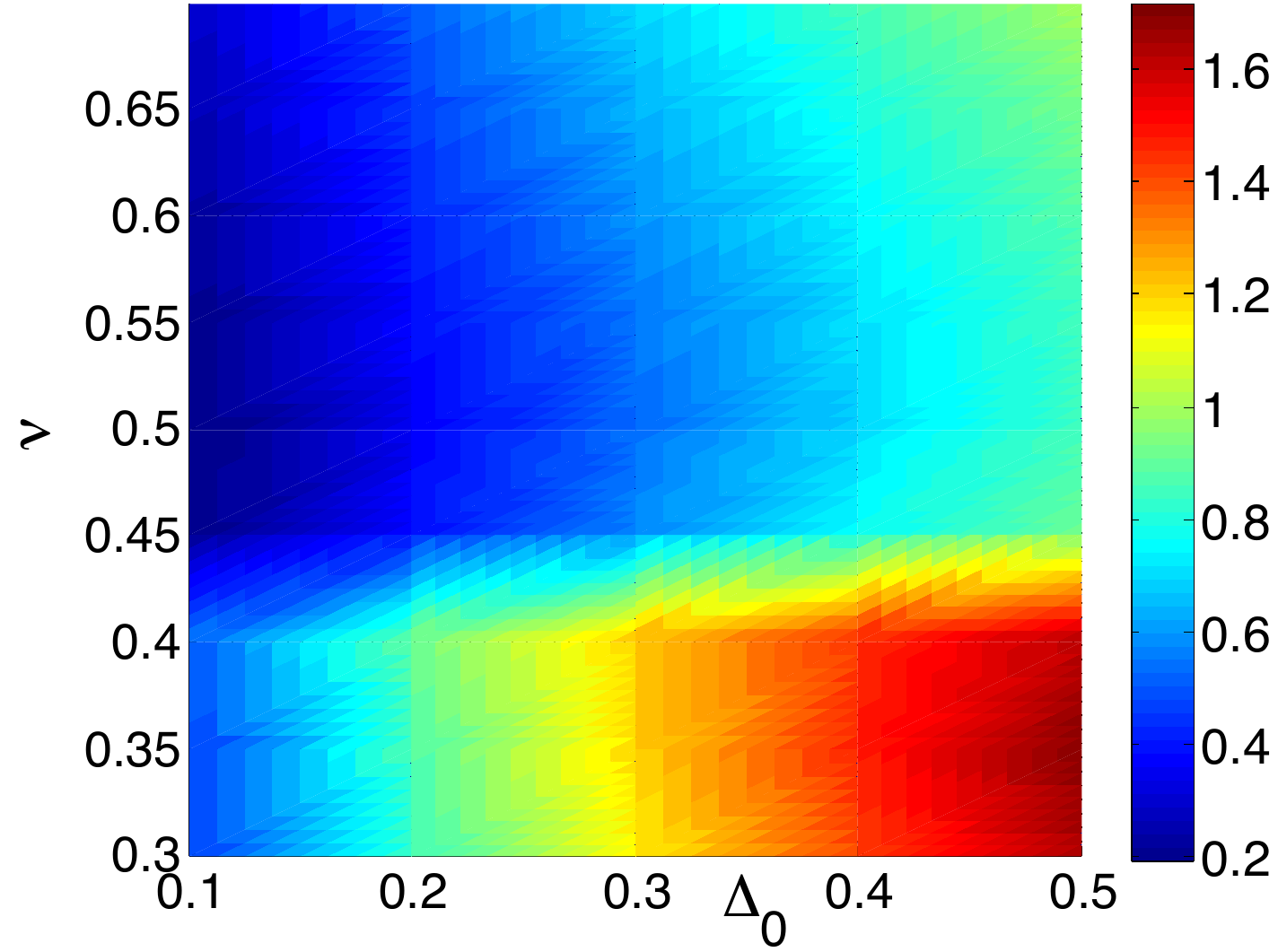}
    \label{ms2}
} \hspace{1cm}
\subfigure[]{
    \includegraphics[width=2.75in]{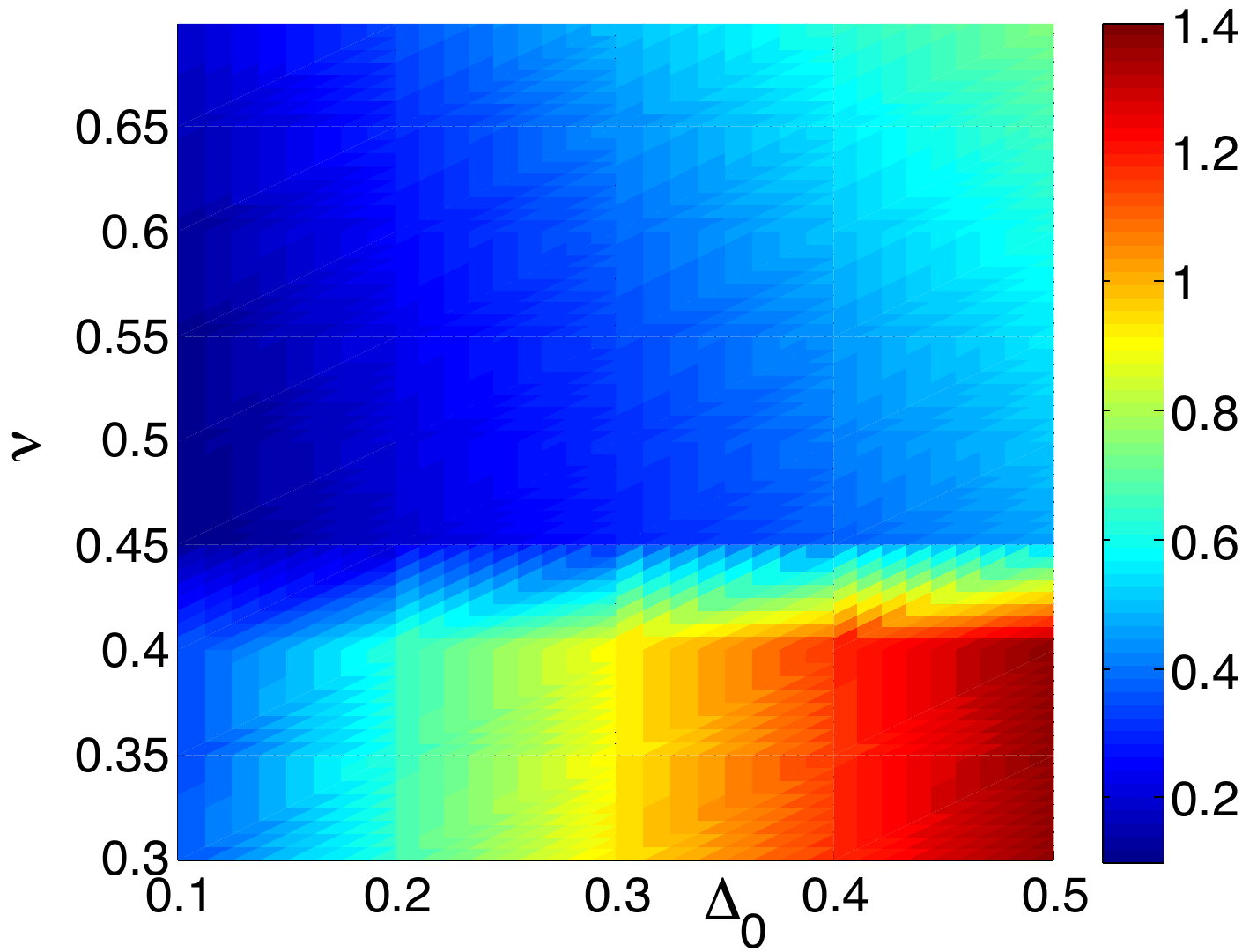}
    \label{ms3}
} \hspace{1cm}
\subfigure[] % caption for subfigure a
{
    \includegraphics[width=2.75in]{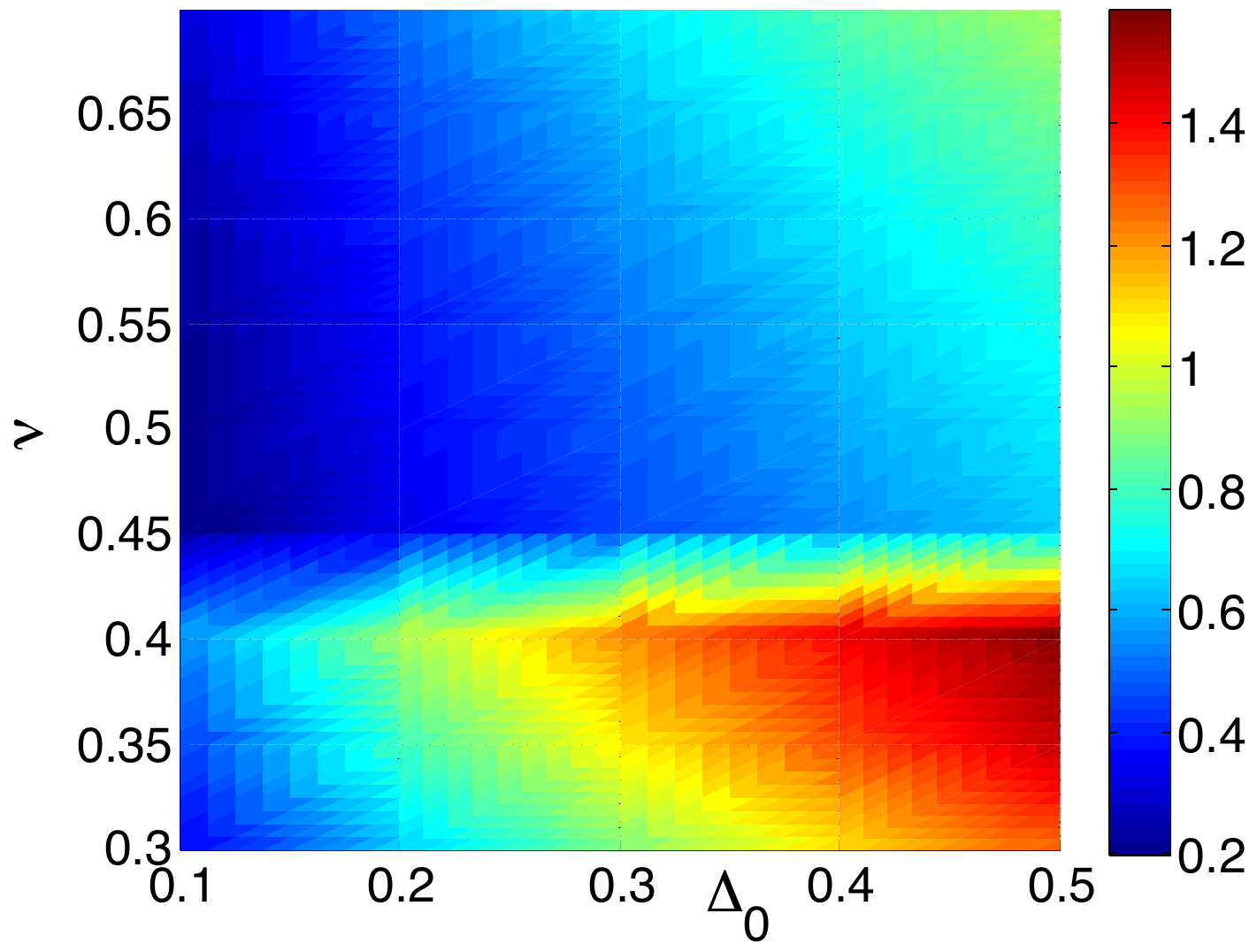}
    \label{ms4}
} %\hspace{1cm}
\caption[]{Gap of the Leggett mode for (a) extended $s$-wave, (b)
  standard $s$-wave, (c)$\Delta = \Delta_0 \sin k_x \sin k_y$, and (d)
  $d$-wave superconducting gaps. The magnitude of $ \Omega_0$ in eV is indicated by the color map to the right of each figure, with blue corresponding to regions of smaller $\Omega_0$, and red to regions of larger $\Omega_0$.
  The vertical axis indicates the filling $\nu$, with $1/2$-filling corresponding to the undoped case;
  the horizontal axis is the scale of the maximum magnitude of the gap in eV: we take $ \Delta = \Delta_0 \Gamma_k$.  }
 \label{Fig_LGComp}
% \end{center}
 \end{figure*}
%\end{widetext}

\section{Results}

Having established the general form of the effective action of the
Leggett mode, we now turn to a quantitative evaluation of the
coefficients in Eq.~(\ref{Eq_SNoC}).  Our principle interest will be
what potential information the Leggett mode can give about the form
of the superconducting gap -- in particular, we address the question of whether it
can distinguish between the popular extended $s$-wave gap and other
plausible pairing symmetries.  Unfortunately, it is clear from our
equations that the Leggett mode properties depend on the absolute
value of the gap function, thereby preventing any qualitative
sensitivity of the mode to a sign-change in the gap function. We
find that the clearest signature is the lifetime of the Leggett mode
as a function of doping -- at low dopings we find the Leggett mode
to lie below the two-particle continuum; at higher doping the mode
is always at higher energies than the two-particle continuum and
hence will give at best a very broad resonance.

\subsection{The Leggett Mode gap} \label{Sect_LeggettGaps}

We begin by studying the Leggett mode gap $\Omega_0$ for several
different gap functions, with the objective of understanding the
qualitative differences expected between these in potential
experiments.  In each case, the mode is expected to be
visible if it lies below the two-particle continuum, which is set by
$2 \min | \Delta|$ (where the minimum is taken over the Brillouin zone).

At $q=0$ and $T=0$, the symmetric and antisymmetric phase
oscillations decouple, and from the effective action
(\ref{Eq_SEffFinal}) the gap of the Leggett mode $ \varphi$ is given
by
 \be \notag
 \Omega_0 = \sqrt{- \frac{M}{N_{\varphi\varphi} } }
 \ee
with $M$ and $N_{\varphi \varphi}$ given by Eqs. (\ref{Eq_Ntot}) and
(\ref{Eq_Mtot0}). Note that $M>0$ and $N_{\varphi \varphi}<0$, so
that the Leggett mode gap is well defined. We can evaluate the
coefficients $N_{\varphi \varphi}$ and $M$ by integrating the
expressions (\ref{Eq_Ntot}) and (\ref{Eq_Mtot0}) numerically over
the Brillouin zone.  We use the values of $\epsilon_{\alpha}$ quoted
in Eq. (\ref{dispersion1}).

Fig. \ref{Fig_LGComp} shows the expected gap of the Leggett
mode for extended $s$-wave, standard $s$-wave, $d$-wave, and $\Delta_0
\sin k_x \sin k_y$ gaps, as a function of the filling fraction $\nu$ and the maximum gap magnitude
$\Delta_0$.  The general form of $\Omega_0$ is
similar in all four cases: it increases with the superconducting gap
$\Delta_0$, and has its lowest values at a filling of approximately
$\nu = 0.4$.  For all four order parameters, we also find
the gap of the Leggett mode shows an academically interesting
chemical potential dependence, droping sharply between $
\nu = 0.4$ and $ \nu=0.5$ independent of the momentum-dependence of the order parameter.

The qualitative features of these plots can be understood by
considering the form of Eqs. (\ref{Eq_Ntot}) and (\ref{Eq_Mtot0}).
First, we see that $\Omega_0$ increases monotonically with
$\Delta_0$, at a slightly less than linear rate. Though naively both
$M$ and $N_{\varphi \varphi}$ scale quadratically with $\Delta_0$,
$N_{\varphi \varphi}$ has divergences if $E_+^{(\Delta)}$ or
$E_-^{(\Delta)}$ vanish; these are cut off by the gap but
nevertheless contribute the major part of the integral.
Consequently, $N_{\varphi \varphi}$ is well approximated by
$N_{\varphi \varphi} \sim V_{FS}/\Delta$, with $V_{FS}$ the volume
of the Fermi surface.  On the other hand,  $M$ vanishes at the Fermi
surface in the limit of small $\Delta$, scaling approximately as $M
~ \Delta$ in this region. Hence the quantity $\sqrt{\frac{M} {-
N_{\varphi \varphi}} }$ increases with $\Delta$, with a power close
to (but slightly less than) $1$.

%\begin{widetext}
\begin{figure*}[ht]
 \begin{center}
\subfigure[]{
    \includegraphics[width=2.75in]{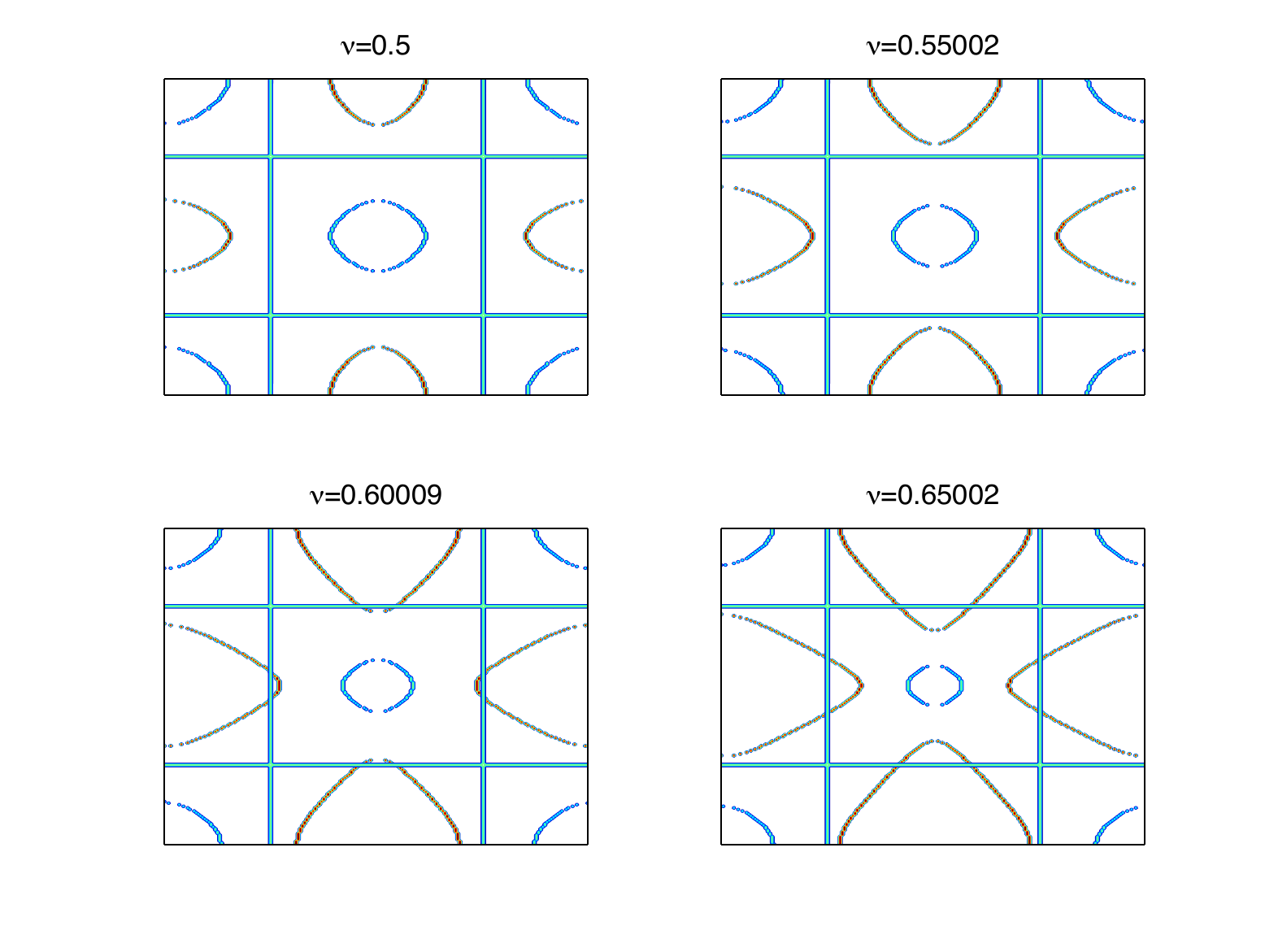}
    \label{Fs1}
} \hspace{1cm}
\subfigure[] % caption for subfigure a
{
    \includegraphics[width=2.75in]{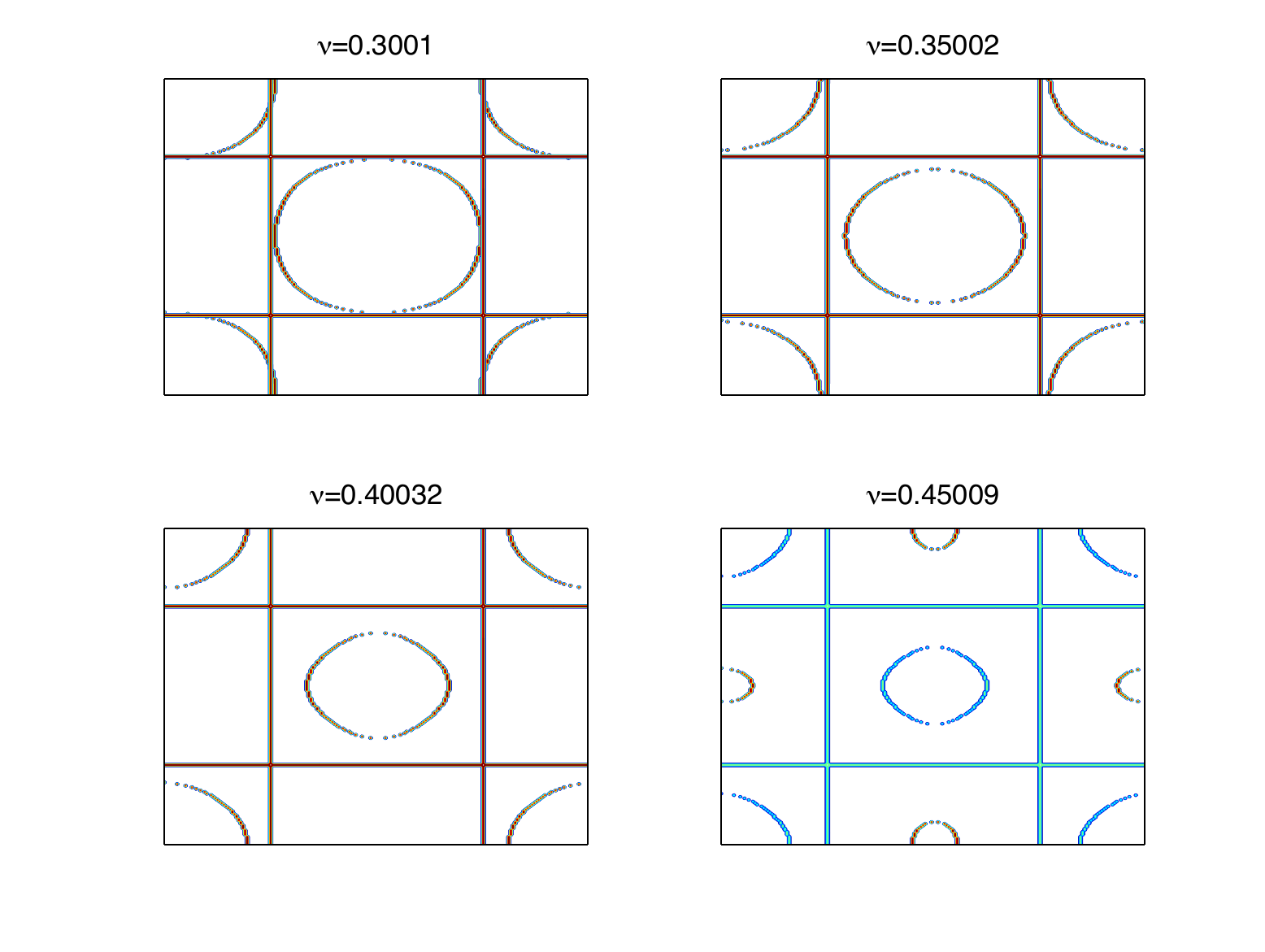}
    \label{Fs2}
} %\hspace{1cm}
\caption{ Plots of Fermi surface as a function of chemical potential
for  $0.3 < \nu < 0.65$.
  The relevant filling fractions are shown as $\nu$ in the title of
  the figure.  The electron pockets first appear at approximately $\nu
  = 0.4$, and cross the nodal lines at $\nu = 0.6$.  The nature of the
  hole pockets does not change substantially over the range shown
  here.}
 \label{FillSeries}
 \end{center}
 \end{figure*}
%\end{widetext}

The non-monotonic dependence on $\nu$, which is similar in all four
cases, stems from the dependence of the shape and volume of the
Fermi surface on the chemical potential. As stated above, the
integral expression for $N_{\varphi \varphi}$ is dominated by
contributions near the Fermi surface.  $M$, on the other hand,
receives significant contributions from the entire Brillouin zone,
and is thus much less sensitive to the shape of the Fermi surface.

To illustrate how these effects play out in our two-orbital model,
Fig.~\ref{FillSeries} plots the Fermi surface of the two-orbital
model in the normal state for the range of dopings considered here.
Between $\nu = 0.4$ and $ \nu = 0.45$, at precisely the locus of the
sharp drop in $\Omega_0$ seen above, a new set of Fermi pockets
appears at the points $(\pm \pi, 0)$ and $(0, \pm \pi)$ as a second
pair of bands crosses the Fermi level in the normal state.  In the
superconducting state this results in more areas where the
expressions for $N_{\varphi \varphi}$ are relatively large -- in
particular, due to the much smaller Fermi velocity near the new
branches of the Fermi surface, the area over which $N_{\varphi
\varphi}$ is large increases sharply, leading to a sudden reduction
in $\Omega_0$. Also worthy of note are the extremal values ($\nu =
0.3, \nu > 0.6$) at which the Fermi surface intersects with the
nodes of the extended $s$-wave gap.    These account for the non-
monotonic behavior of $\Omega_0$ observed in both extended $s$-wave
and $d$-wave order parameters between $\nu = 0.3$ and $0.4$, as the
cut-off in the normal-state divergences of $N_{\varphi \varphi}$
grows smaller. Though the application of our simple two-orbital
model at large fillings is not warranted, and the features discussed
in this paragraph are model-dependent, we expect them to be accurate
for gaps diagonal in the orbital basis inasmuch as the band
structure given by the two-orbital model is correct.

\subsection{Observability of the Leggett mode}

\begin{figure*}[htp]
 \begin{center}
 \subfigure[]{
    \includegraphics[width=2.75in]{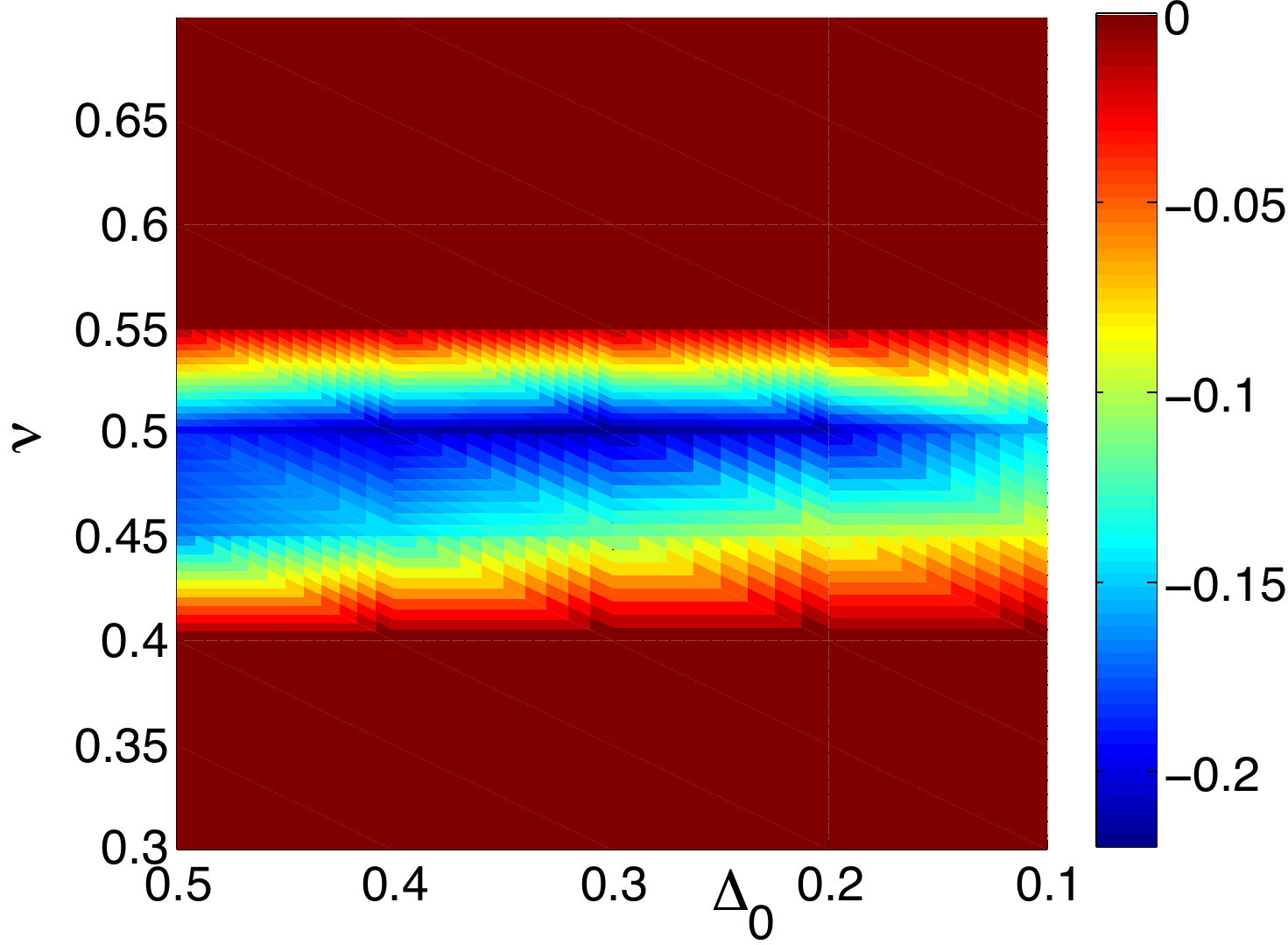}
    \label{Swave1}
} \hspace{1cm}
\subfigure[] % caption for subfigure a
{
    \includegraphics[width=2.75in]{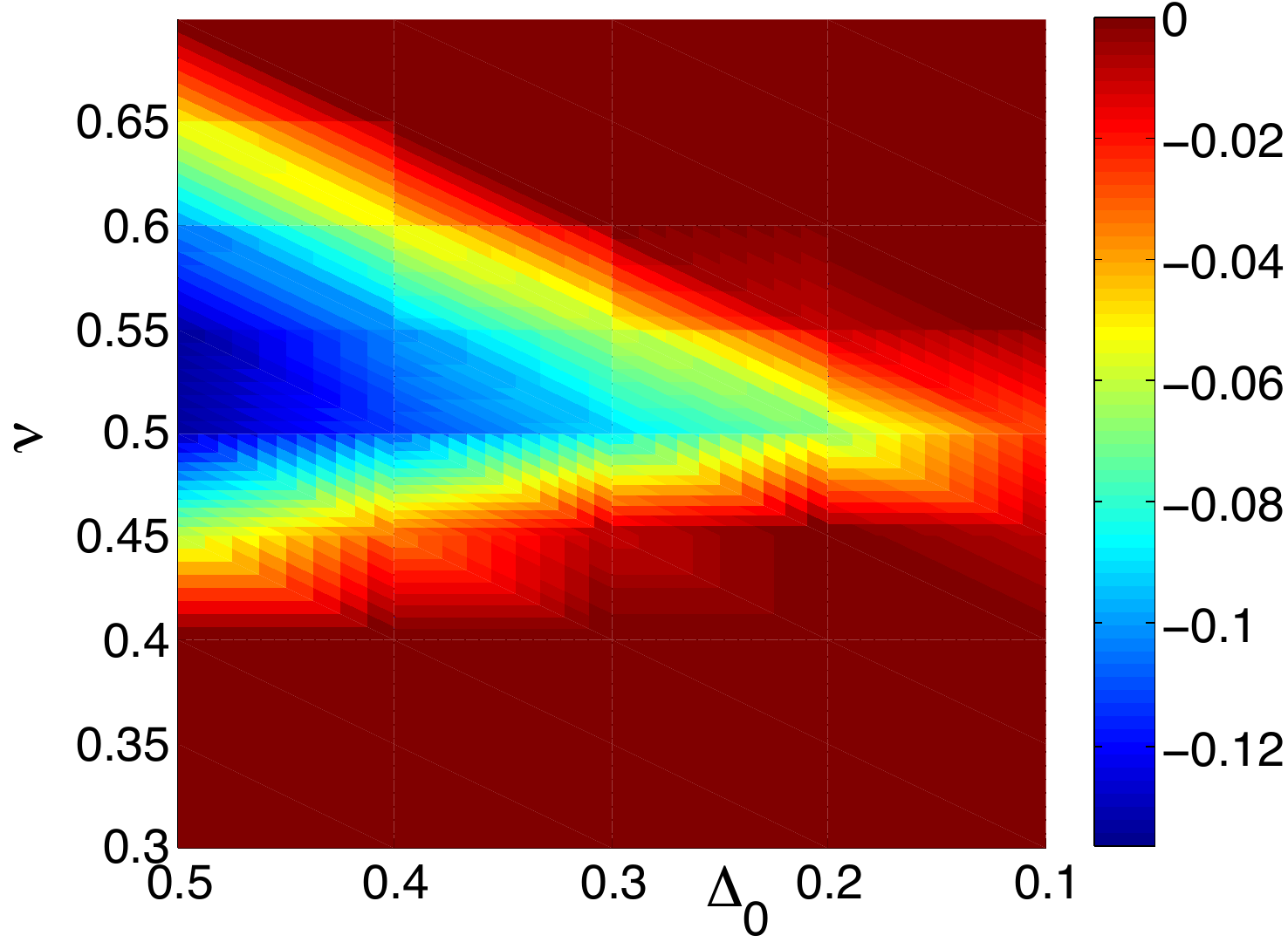}
    \label{Swave2}
} %\hspace{1cm}
\caption[]{ Distance between the Leggett mode and two-particle
continuum for (a) extended $s$-wave and (b) ordinary $s$-wave  gaps,
as a function of filling fraction $\nu$ and gap magnitude $\Delta_0$
in eV.  Dark maroon area indicates regions where the Leggett mode
lies above the two-particle continuum.  Blue regions indicate
maximum distance from the two-particle continuum.  The vertical
colorbar indicates the energy scale of these differences, as a
fraction of the two-particle continuum $2 \min \Delta_0$.
  }
 \label{Fig_LGComp2}
 \end{center}
 \end{figure*}

In order for the Leggett mode to give a sharp resonance in
experiments, it should lie below the two-particle continuum. For the
$d$-wave and sine-wave gaps, which are nodal for the iron-based
superconductors' Fermi surfaces, this is obviously never the case.
For ordinary $s$-wave and extended $s$-wave gaps, the position of
the Leggett mode at $q=0$ relative to the two-particle continuum
depends on the values of $\nu$ and $\Delta_0$. Figure
\ref{Fig_LGComp2} plots distance between the gap of the Leggett mode
and the minimum energy of the two-particle continuum as a function
of $\Delta_0$ for both extended $s$-wave ($\Gamma_k = \cos k_x \cos
k_y $) and standard $s$-wave $(\Gamma_k =1$) gaps.

%\begin{widetext}
%\end{widetext}

The principle difference between the two nodeless gaps is the range
of dopings over which the Leggett mode is expected to be observable.
In the pure $s$-wave case, the two-particle continuum is given by $2
\Delta_0$, independent of the shape of the Fermi surface.  Hence, as
seen in Figure \ref{Swave2}, the dominant effect here is that the
gap of the Leggett mode scales sublinearly in $\Delta_0$,  and hence
becomes observable only at large values of the gap.  Its separation
from the two-particle continuum is extremely small at the small
values of $\Delta_0$ expected to occur near half-filling. (Away from
half-filling, the gap of the Leggett mode lies above the
two-particle continuum, as shown in the figure, due to the
Fermi-surface effects discussed above). In the extended $s$-wave
case, however, the minimum of the gap also depends on how close the
Fermi surface comes to the nodes of the gap function.  Hence the
dependence on filling fraction here is more pronounced (Figure
\ref{Swave1}); for all values of $\Delta_0$ the Leggett mode sits
definitively below the two-particle continuum in the interval $0.45
< \nu < 0.5$. Further from half-filling, where the nodes of the
extended $s$-wave gap sit closer to the Fermi surface, the mode is
never visible.   $\Omega_0$ is smaller overall in the extended $s$-
wave case, compensating for the fact that the minimum of the
two-particle continuum is de facto smaller than in the standard
$s$-wave case.  Most notably, for the small values of $\Delta_0$
expected near $1/2$ -filling, we expect the Leggett mode to be below
the two-particle continuum in the extended $s$-wave case

\subsection{Dispersion of the Leggett mode for extended $s$-wave gap}

We now return to the general form of the effective action for the
phase degrees of freedom, and analyze the structure of its modes at
small $q$.  In the absence of the Coulomb interaction, the form of the dispersion
is effectively characterized by
\begin{widetext}
\be
 w^2 = \frac{1}{2} \left [  q^2 ( c_{\phi \phi}^2 + c_{\varphi
    \varphi}^2) + \Omega_0^2 \pm \sqrt{
 \left \{ q^2( c_{\varphi \varphi}^2-c_{\phi \phi}^2  )+ \Omega_0^2
 \right \}^2 +
    \frac{4 c_{\phi \varphi}^4}{N_{\phi \phi} N_{\varphi \varphi}} q^4 \cos^2 2
    \theta} \right ] \ \ \ ,
\ee
where $\theta$ is the angle in the $(k_x, k_y)$ plane.
If $c_{\phi \varphi}$ vanishes, we retrieve the gapless Goldstone
mode, and the gapped Leggett mode.  From Eq. (\ref{Eq_SEffFinal}),
we find that adding the Coulomb term modifies
this according to:
\begin{align}\notag
w^2 = & \frac{1}{2} \left [  \left \{ c_{\phi \phi}^2 (1 -4 N_{\phi
      \phi} U(q)) + c_{\varphi \varphi}^2 \right \} q^2 +
     \Omega_0^2 \right] \\
     &\pm \frac{1}{2} \sqrt{  \left \{ q^2 (c_{\varphi \varphi}^2 -
         c_{\phi \phi}^2 (1 - 4 N_{\phi \phi}U(q) )) +
          \Omega_0^2 \right \}^2 + \frac{4 c_{\phi \varphi}^4 q^4 (1-4
          N_{\phi \phi}
          U(q) )}{N_{\phi \phi} N_{\varphi \varphi} }  \cos^2 ( 2
        \theta ) } \ \ \ .
\end{align}
\end{widetext}
In this case, taking the negative sign for $U(q) \sim q^{-1}$, (the
unscreened Coulomb interaction in $2$D) results in $\omega^2 < 0$,
indicating that the Goldstone mode has been replaced by a plasma
mode.  The structure of the Leggett mode is, however, largely
unchanged by the presence of the Coulomb interaction.  In
particular, we still have $\lim_{q\rightarrow -} \omega(q) =
\Omega_0$.

\begin{figure}[ht]
 \begin{center}
\subfigure[]{
    \includegraphics[width=2.75in]{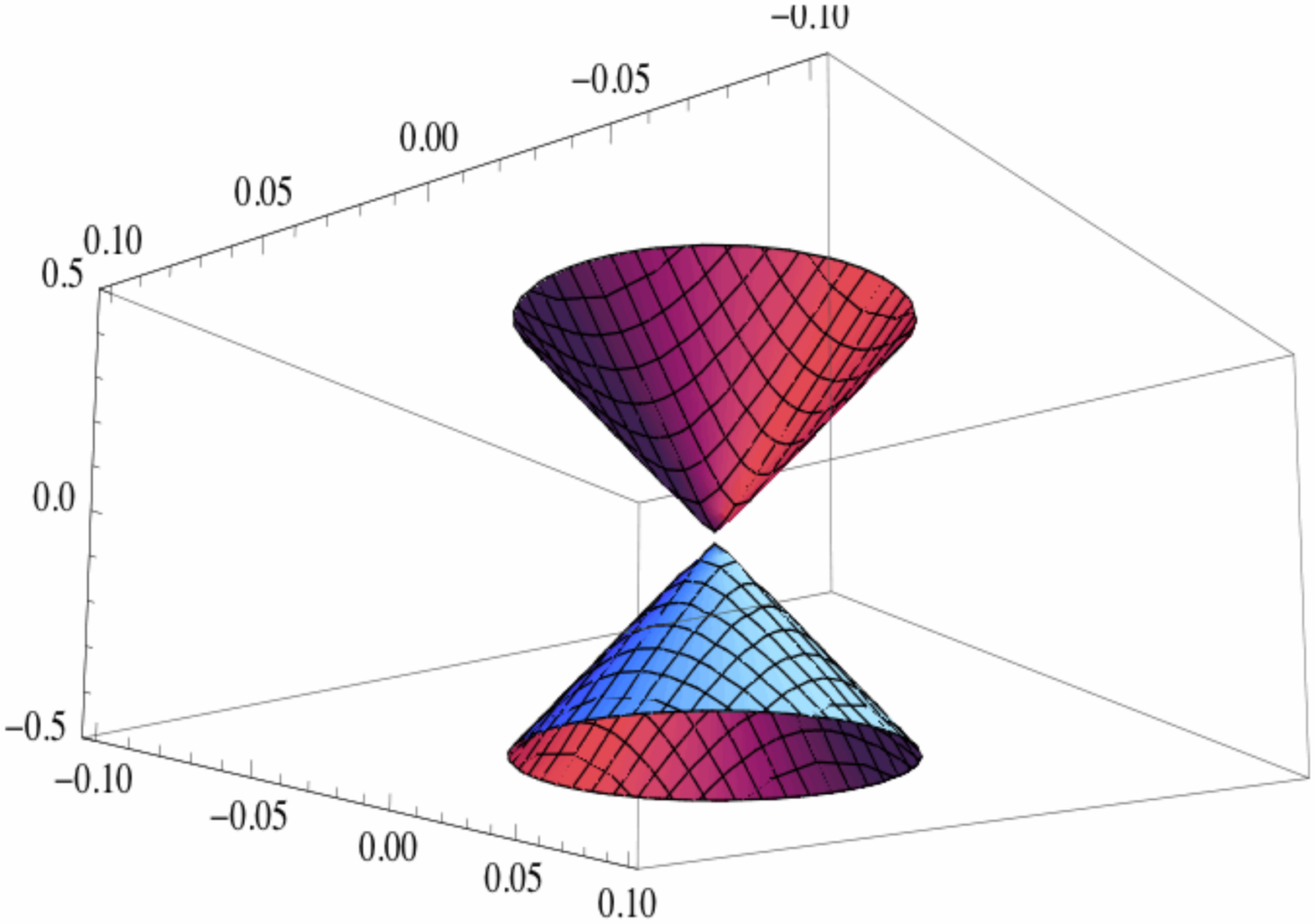}
    \label{Fig_disp}
} \hspace{1cm}
\subfigure[] % caption for subfigure a
{
    \includegraphics[width=2.75in]{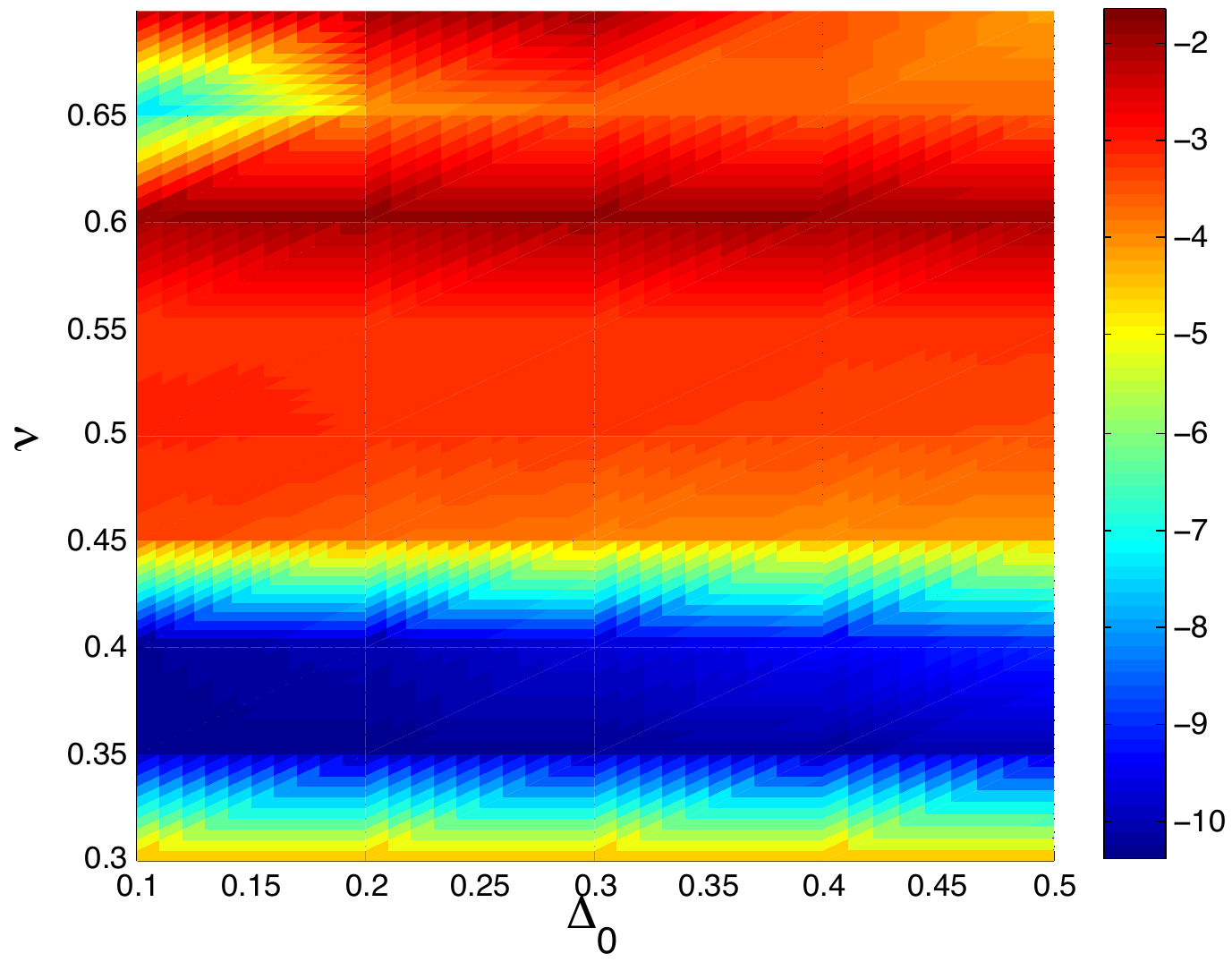}
   \label{Fig_Vprofs}
}
\caption{ Velocity of the dispersion relations $\Omega$ vs $q$ as a function of filling $\nu$ and gap magnitude $\Delta_0$ for the extended $s$-wave gap $\Delta = \Delta_0 \cos k_x \cos k_y$.  For small $q$, the qualitative shape of the
  dispersion (a) (shown at $\Delta_0 = 0.1, \nu =0.4 $ ) does not depend strongly on either the form of $U$ or the precise choice of $\Delta_0$ and $\nu$.   However, the velocity, plotted as a function of  $\Delta_0$ and $\nu$ in (b), is
  sharply sensitive to the Leggett gap and chemical potential.   In particular, the magnitude of the veolcity increases sharply as the filling fraction decreases. }
 \end{center}
 \end{figure}

Figure \ref{Fig_disp} plots the dispersion relation $\Omega(k)$ for
the low-energy modes for several values of $\Delta_0, \mu$.  As we
have kept only terms to quadratic order in $q, \Omega$, we expect
this to be valid for small $q$, and have restricted the range of the
plots accordingly.  The important feature to note is that the
velocity anisotropy, due to the off-diagonal terms $c_{\phi \varphi}
(q_x^2-q_y^2)$, is relatively small and the dispersion is
approximately rotationally invariant. Because of this, the
dispersion relation of the Leggett mode is well-characterized by the
gap $\Omega_0$ (c.f. Sect. \ref{Sect_LeggettGaps}), and the velocity
$v \equiv \lim _{q \rightarrow \infty} \omega(q)$.  This latter is
plotted for the extended $s$-wave gap in Fig. \ref{Fig_Vprofs}.

\section{Conclusions}
We have obtained the fluctuation action for the superconducting
phase collective modes of a two-orbital model for iron-based
superconductors, with particular emphasis on the antisymmetric
Leggett mode. By fixing the parameters of the band structure, our
calculation has identified the range of doping and superconducting
gap magnitude over which the undamped Leggett mode exists below the
two-particle continuum.  As the Leggett mode's visibility increases 
with the magnitude of the superconducting gap, this result also suggests that
if the bandwidth is narrower, there is a higher
possibility of observing the undamped Leggett mode. Therefore, a
strong renormalization of bands could enhance the existence of an
undamped Leggett mode. Unfortunately, the mode and its dispersion
are insensitive to the sign of the order parameter on Fermi
surfaces, and the mode does not qualitatively  distinguish
between the sign-changed $s$-wave and a normal $s$-wave
superconductors. However, we find that quantitative characteristics
of the mode can in principle distinguish between such different
pairing symmetries.  First, we find that the Leggett mode does lie below the
two-particle continuum, near half filling and sufficiently deep in
the superconducting region, for the extended as well as normal
$s$-wave gap.  This is distinct from the case of nodal gaps, where
low-energy quasi-particles are always expected to broaden the
Leggett mode resonance.  Second, we find that the difference in signatures
between the two kinds of $s$-wave pairing symmetry investigated here
is subtle, but that the extended $s$-wave gap is visible over a
narrower range in doping, but further below the two-particle
continuum over much of its range of detectability. This
difference
comes from the different doping dependence of the two $s$ wave gap
functions: unlike the normal, sign-unchanging $s$-wave gap, the
$s^\pm$ order parameter will most likely change upon doping as the
Fermi surfaces become closer to the line of zeroes that a
sign-changing gap should have in the Brillouin zone. This gap
variation upon doping is present in both strong and weak coupling
models \cite{thomalefrg2009}. In this situation, the Leggett mode
will move from a relatively sharp mode below the two-particle
continuum into a strongly damped  mode above the two-particle
continuum as doping is increased. This quantitative change can in
principle be observed in experiments.

It has been claimed that the Leggett mode has been observed in
MgB$_2$ by Raman scattering\cite{Mgb2} and point-contact transport
measurements\cite{Mgb3}, although the energies of the Leggett mode
measured in the two experiments are different.  Ref. 
\onlinecite{chubukov2009} found that in the weak-coupling treatment 
of superconductors with an $s_\pm$ gap, however, the $A_{1g}$
Leggett mode does not couple to Raman scattering. 
The analysis carried out here relies heavily on the fact that
 iron-based superconductors are more strongly coupled than MgB$_2$,
and that the superconducting phase is thus well-described by 
considering the orbital, rather than the band, basis.  
This leads to a result which differs from that of the weakly-coupled 
approach in two ways.  First, the strong-coupling approach suggests that
the Leggett mode should be observable 
in Raman spectra.  Second, in the strong-coupling treatment,
the different orbital symmetries should be kept explicitly when 
determining the relevant Raman channels. For our
model, the Leggett mode is caused by an oscillation between the
condensates involving the scattering of a pair of $d_{xz}$-orbital
electrons into a pair of $d_{yz}$-orbital electrons. Such a process
causes a relative density fluctuation, $\delta n=n_{xz}-n_{yz}$,
between two orbitals, which belongs to the $B_{1g}$ irreducible
representation of the point group ($D_{4h}$) of the crystals.
Therefore, the Leggett mode is a $B_{1g}$ mode in this orbital-based
model, and should exist in the $B_{1g}$ channel in
 Raman scattering experiments.  
(Without the orbital characters, the Leggett mode should be a
pure $A_{1g}$ mode, as is the case  in MgB$_2$~\cite{Mgb2}).  Thus, 
observing the Leggett mode in channels other than $A_{1g}$ 
should provide important evidence about the orbital structure of
condensed pairs in the iron-based superconductors.

\section{Acknowledgements}
We wish to thank  X. Dai, F. Wang, D.-H. Lee, A. Chubukov and I. Eremin, for comments
and suggestions. BAB acknowledges support  from the Alfred P. Sloan
Foundation and from Princeton University startup funds. BAB wishes to acknowledge the hospitality of the Institute of Physics in Beijing for the period during which the current work was finalized. MMP
acknowledges support from the EPSRC.  FJB acknowledges the hospitality of KITP.

\begin{appendix}

\begin{widetext}

\section{Calculating the phase-only effective action}\label{DerivationApp}

We begin by deriving the effective phase-only action for a generic
Hamiltonian of the form:
 \ba
 H = \sum_{\alpha, \beta, r, r'} \sum_{\sigma = \uparrow, \downarrow} c^\dag_{\alpha, r, \sigma}
   (\epsilon(r, r')_{\alpha \beta}  - \mu\delta_{\alpha \beta})  c_{\beta,
     r', \sigma}+  V_{\alpha, \beta}(r; r') b^\dag_{\alpha, r}
b_{\beta, r'}
 +i e
U (r-r')\rho_{ r} \rho_{ r'}
 \ea
Here $\alpha, \beta$ are orbital indices, $V$ is the superconducting
interaction (in our model, a spin-spin antiferromagnetic  interaction decoupled in
the Cooper channel to give us $\cos(k_x) \cos(k_y)$ pairing) , and
$U$ is the Coulomb interaction. The density is given by $\rho_{r} =
\sum_{\alpha \sigma }c^\dag_{\alpha,r, \sigma} c_{\alpha, r,
\sigma}$, and the superconducting bilinear is $b^\dag_{\alpha, r} =
c^\dag_{\alpha, r, \uparrow} c^\dag_{\alpha, r, \downarrow}$.

We may decouple the two $4$-fermion interactions by means of two
Hubbard-Stratonovich transformations.  This decoupling gives the action:
 \ba \label{Eq_FProp}
S  &=& \sum_{\alpha, \beta, r, r'} \left \{
 \sum_{ \sigma}
 c^\dag_{\alpha, r, \sigma} \left [ (\partial/ \partial_\tau  -
 \mu ) \delta_{\alpha, \beta}+ \epsilon_{\alpha, \beta} (r, r') \right
 ]c_{\beta, r', \sigma} - \delta_{r, r'} \left [
 \Phi_{\alpha r}  b_{\alpha, r'} +\Phi^\dag_{\alpha r}
 b^\dag_{\alpha, r'} \right ]
\right . \n
 && \left .  -i e \delta_{r,r'} \left [  \chi_{r}  \rho_{r'} +
    \chi^*_{ r}  \rho_{r'}   \right ]
 - \Phi^\dag_{ \alpha r} V^{-1}_{\alpha, \beta}(r; r') \Phi_{\beta r'} -
 \chi_{ r } U (r- r')^{-1} \chi_{ r'} \right \}
 \ea
where $\Phi$ is the Hubbard-Stratonovich field associated with the
superconducting interaction, and $\chi$ is associated with the
Coulomb interaction.

In computing this effective action, we follow the method of ref.
\onlinecite{SharapovBeck} to isolate the action for the phase
degrees of freedom.  That is, taking $\Phi_{\alpha r}
=\Delta_{\alpha r} e^{i \theta_{\alpha}(r)}$, we perform the gauge
transformation \be c_{\alpha, r, \sigma} \rightarrow e^{i
\theta_\alpha(r)/2 } c_{\alpha,
  r, \sigma}
\ee
to absorb all terms involving the phase of the superconducting order
parameter into the first term of Eq. (\ref{Eq_FProp}).  We then have:
\ba \label{Eq_Prop2}
S  &=& \sum_{\alpha, \beta, r, r'} \left \{
 \sum_{ \sigma}
 c^\dag_{\alpha, r, \sigma} e^{-i \theta_\alpha(r)/2} \left [ (\partial/ \partial_\tau  -
 \mu ) \delta_{\alpha, \beta}+ \epsilon_{\alpha, \beta} (r, r') \right
 ] c_{\beta, r', \sigma} e^{i \theta_\alpha(r')/2}  - \delta_{r, r'} \left [
 \Delta_{\alpha r} \left( b_{\alpha, r'} +
 b^\dag_{\alpha, r'} \right ) \right ]
\right . \n
 && \left .  -i e \delta_{r,r'} \left [  \chi_{r}  \rho_{r'} +
    \chi^*_{ r}  \rho_{r'}   \right ]
 - \Phi^\dag_{ \alpha r} V^{-1}_{\alpha, \beta}(r; r') \Phi_{\beta r'} -
 \chi_{ r } U (r- r')^{-1} \chi_{ r'} \right \}
\ea

It is convenient to re-express the kinetic terms as:
\ba \label{Eq_BCSBasis}
&\sum_{\alpha, \beta, r, r'} \left \{ \right . &
 c^\dag_{\alpha, r, \uparrow} e^{-i \theta_\alpha(r)/2} \left [ (\partial/ \partial_\tau  -
 \mu ) \delta_{\alpha, \beta}+ \epsilon_{\alpha, \beta} (r, r') \right
 ]c_{\beta, r', \uparrow} e^{i \theta_\beta(r')/2} \n
&&\left.  -  c_{\alpha, r, \downarrow} e^{-i \theta_\beta(r')/2} \left [ (-\partial/ \partial_\tau  -
 \mu ) \delta_{\alpha, \beta}+ \epsilon_{\beta, \alpha} (r', r) \right
 ] c^\dag_{\beta, r', \downarrow} e^{i \theta_\alpha(r)/2}  \right .\n
&& \left.
+\delta_{\alpha, \beta}  \delta_{r, r'} e^{-i \theta_\alpha(r)/2} \left [ (\partial/ \partial_\tau  -
 \mu ) + \epsilon_{\alpha, \beta} (r', r) \right
 ]  e^{i \theta_\beta(r')/2} \right \}
\ea
The first two terms can now be combined with the rest of the fermionic
action can be expressed in matrix form in the BCS basis. The
final line of Eq. (\ref{Eq_BCSBasis}) gives a separate contribution to
the action, which can be expressed, to quadratic order in
$\theta, \partial_r \theta$:
\be \label{Eq_Extras}
\int d^2 r d \tau \left \{ \frac{i}{2} \frac{\partial}{\partial_\tau}
  \theta_\beta(r)  + \int d^2k \left( \frac{1}{2} \frac{\partial}{ \partial_{r_i}}\theta_\beta(r)  \frac{\partial}{\partial_{k_i}} \epsilon_{\beta, \beta}
- \frac{1}{8}\theta_\beta(r)  \frac{\partial^2}{\partial_{ r_i} \partial_{r_j}}
\theta_\beta(r) \frac{\partial^2}{\partial_{k_i} \partial_{k_j}}\epsilon_{\beta, \beta} (k) \right )  \right \}
\ee
The first term is a total derivative and will not contrubute to the dynamics of the phase-only effective action.   The second term in any case vanishes, as $v$ is odd over the Brillouin zone.  Hence only the last term appears in the effective action.

As we are interested in the dynamics of the phase degrees of
freedom, we replace $\Delta_{\alpha, r} $ by its mean-field value.
For the time being, we drop the Coulomb terms by setting $U =0, \chi
=0$; we will discuss these further in Sect. \ref{CoulombApp}.  We
further consider only the slowly varying phase fluctuations.  This
allows us to expand the exponentials of the fermionic terms in
(\ref{Eq_Prop2}) using: \ba e^{-\frac{i}{2}( \theta_\alpha(r)-
\theta_\alpha(r') ) } \approx 1 -\frac{i}{2}( \theta_\alpha(r)-
\theta_\alpha(r') ) -\frac{1}{4}( \theta_\alpha(r)-
\theta_\alpha(r') )^2  \n e^{-\frac{i}{2}( \theta_1(r)- \theta_2(r')
) } \approx e^{-\frac{i}{2}
  \varphi_0 } \left \{ 1
-\frac{i}{2}( \theta_1(r)- \theta_2(r') - \varphi_0) -\frac{1}{4}(
\theta_\alpha(r)- \theta_\alpha(r') - \varphi_0 )^2 \right \}
\ea
where we have explicitly separated out the possible background
expectation value of the phase difference $\varphi_0$ between the
gaps.  In practice $\varphi_0=0$ is set by the mean-field equations.

Defining $\psi^\dag_{r} = ( c^\dag_{1, r, \uparrow}, c_{1, r ,
 \downarrow}, c^\dag_{2, r, \uparrow}, c_{2, r, \downarrow} )$, we may
now express the first two terms in Eq. (\ref{Eq_BCSBasis}), after Fourier
transforming, as
\be \label{Eq_S}
S_{Fermi} = \int d\omega_1 d\omega_2 \frac{d^2 k_1}{4 \pi^2} \frac{d^2 k_2}{4 \pi^2}
\psi^\dag_{k_1, \omega_1}  G^{-1}(k_1, k_2, i\omega_1, i\omega_2)
\psi_{k_2, \omega_2}
\ee
with \be \label{Eq_ForG}
 G^{-1}(k_1, k_2, i\omega_1, i\omega_2) =   G_0^{-1}(k_1,\omega_1)
\delta_{q,0}\delta_{\Omega,0} + \Sigma(k_1, k_2, i \omega_1, i\omega_2)
 \ee
where here $q \equiv k_1-k_2, \Omega\equiv \omega_1-\omega_2$.  We have:
 \ba
G_0^{-1}(k, \omega) =  \left ( \begin{array}{cccc} i \omega+ \epsilon_{x} & \Delta_1&
 \epsilon_{xy}& 0 \\
\Delta_1 & i \omega-\epsilon_{x} & 0 & -\epsilon_{xy}\\
\epsilon_{xy}& 0 & i \omega+\epsilon_{y}& \Delta_2 \\
0 & -\epsilon_{xy} & \Delta_2 & i \omega- \epsilon_{y} \\
 \end{array} \right)
\ea
where $\epsilon_\alpha \equiv \epsilon_\alpha (k) $ is the kinetic
energy in the orbital basis,
and $\Delta_\alpha \equiv \Delta_\alpha (k)$ is the momentum-dependend
superconducting gap in each orbital.  The second part of
(\ref{Eq_ForG}) is given by:
\begin{eqnarray} \label{Eq_ForSigma}
\Sigma(k, q, i \omega, i \Omega)& =& - \frac{\Omega}{2} \left(
                                                                           \begin{array}{cc}
                                                                             \theta_1\sigma_z & 0 \\
                                                                             0 & \theta_2 \sigma_z \\
                                                                           \end{array}
                                                                         \right)
                                                                         +
                                                                         \frac{i}{2}
                                                                         \left(
                                                                                      \begin{array}{cc}
                                                                                        \theta_1\delta_q( \epsilon_{x}) \sigma_0&
                                                                                        \left
                                                                                         [\theta_2
                                                                                        \epsilon_{xy}(k_1)
                                                                                        -
                                                                                        \theta_1
                                                                                        \epsilon_{xy}(k_2)
                                                                                      \right
                                                                                      ]
                                                                                        \sigma_0
                                                                                        \\
                                                                                        \left
                                                                                         [\theta_1
                                                                                        \epsilon_{xy}(k_1)
                                                                                        -
                                                                                        \theta_2
                                                                                        \epsilon_{xy}(k_2)
                                                                                      \right
                                                                                      ]
                                                                                        \sigma_0
                                                                                        &\theta_2
                                                                                        \delta_q(
                                                                                        \epsilon_{y})
                                                                                        \sigma_0\\
                                                                                      \end{array}
                                                                                    \right)
                                                                                    \n
&& - \frac{1}{8} \sum_{k_3, i\omega_3}\left(\begin{array}{cc} \theta_1(k_3, i\omega_3) & \theta_2(q- k_3, i(\Omega-\omega_3)) \end{array} \right)
     \left(\begin{array}{cc}  \delta^{(2)}_{q, k_3}  (\epsilon_{x}) \sigma_z & 0 \\
       0 & \delta^{(2)}_{q, k_3}  (\epsilon_{x}) \sigma_z \\
        \end{array}  \right)
   \left(\begin{array}{c} \theta_1(q- k_3, i(\Omega-\omega_3))\\
       \theta_2(q- k_3, i(\Omega-\omega_3)) \end{array} \right)  \n
&&
- \frac{1}{8}  \sum_{k_3, i\omega_3}\ \left(\begin{array}{cc}  0 & B(k_1, i \omega_1, k_2,
  i\omega_2, k_3. i\omega_3) \sigma_z \\
       C(k_1, i \omega_1, k_2,
  i\omega_2, k_3. i\omega_3)\sigma_z & 0 \\
        \end{array}  \right)
\end{eqnarray}
where $\theta_\alpha \equiv \theta_\alpha (q, i\Omega)$, and we have
defined the discrete derivatives:
\begin{eqnarray}
\delta_q \epsilon_{\alpha \beta} &=&  \epsilon_{\alpha \beta}(k_1) -
\epsilon_{\alpha \beta}(k_2)) \n
\delta^{(2)}_q  \epsilon_{\alpha \beta} &=& \epsilon_{\alpha
  \beta}(k_1)-\epsilon_{\alpha \beta}(k_2+k_3)- \epsilon_{\alpha
  \beta}(k_1- k_3)+\epsilon_{\alpha \beta}(k_2)
\end{eqnarray}
The off-diagonal terms quadratic in the phases are:
\begin{eqnarray}  \label{Eq_ABCs}
B(k_1, i \omega_1, k_2, i \omega_2, k_3, i \omega_3)&=& (\theta_2(k_3,
i\omega_3)\theta_2(q- k_3, i(\Omega-\omega_3))
\epsilon_{xy}(k_1)-\theta_2(k_3, i\omega_3)\theta_1(q- k_3,
i(\Omega-\omega_3))\epsilon_{xy}(k_2+k_3) \n
& &- \theta_1(k_3, i\omega_3)\theta_2(q- k_3,
i(\Omega-\omega_3))\epsilon_{xy}(k_1- k_3)+\theta_1(k_3,
i\omega_3)\theta_1(q- k_3, i(\Omega-\omega_3))\epsilon_{xy}(k_2)) \n
C(k_1, i \omega_1, k_2, i \omega_2, k_3, i \omega_3)&=& (\theta_1(k_3,
i\omega_3)\theta_1(q- k_3, i(\Omega-\omega_3))
\epsilon_{21}(k_1)-\theta_1(k_3, i\omega_3)\theta_2(q- k_3,
i(\Omega-\omega_3))\epsilon_{xy}(k_2+k_3) \nonumber \\
&&- \theta_2(k_3, i\omega_3)\theta_1(q- k_3, i(\Omega-\omega_3))\epsilon_{xy}(k_1- k_3)+\theta_2(k_3, i\omega_3)\theta_2(q- k_3, i(\Omega-\omega_3))\epsilon_{xy}(k_2)) \n
\end{eqnarray}

Thus in our treatment, the block diagonal terms
 involve only discrete differences of the band energies, which will become
 derivatives when the momentum of the phase variables is small.  The
 off-diagonal terms contribute, as well as such differences, a term
 which is finite at $q=0$ (or $k_1 =k_2$).  Hence the gap of the
 Leggett mode is, in the absence of inter-orbital pairing, generated
 by the kinetic mixing between the two orbitals.

To obtain the effective action, we integrate out the fermions in
Eq. (\ref{Eq_S}).   In practice, we must evaluate the result
perturbatively in $\Sigma$.  Specifically, we have:
\ba \label{Eq_S2}
S_{eff} &=& S_{MF} - Tr \ln (1-G_0 \Sigma) \n
&\approx& S_{MF} + Tr \left( G_0 \Sigma\right)+ \frac{1}{2 }Tr \left(
  G_0 \Sigma G_0 \Sigma \right)
\ea
where $S_{MF}$ is the mean-field action, from which we
self-consistently determine the values of $\Delta_1, \Delta_2$.  Here
we will evaluate the low-energy, long-wavelength limit of the
effective action (\ref{Eq_S2}) by keeping terms to quadratic order in
$q, \Omega$, and $\theta_\alpha(q,\Omega)$.

\subsection{Evaluating $ Tr G_0 \Sigma$ and $Tr G_0 \Sigma G_0 \Sigma$} \label{TraceApp}

For reference, here we give a more detailed account of the calculation
in Sect. \ref{CalcSect}.

The separate expressions for the two traces are:
\ba \label{M1}
Tr(G_0 \Sigma) &=&  \int \frac{d^2 k}{ (2 \pi)^2} \left
  \{-\frac{\epsilon_{xy}}{4}\left(\varphi(q, i \Omega) \varphi (-q,
    -i \Omega) \right ) \left [1 -
    \frac{\epsilon_{xy}}{(E_+-E_-)E_+^{(\Delta)}E_-^{(\Delta)}}
    \left( E_-^{(\Delta)}E_+ -E_+^{(\Delta)}E_- \right)  \right ]
\right \} \n
&&-  \int \frac{d^2 k}{ (2 \pi)^2} \left
  \{ \frac{q_i q_j}{8}\frac{}{} \left ( \begin{array}{cc} \theta_1(q)&
    \theta_2 (q) \end{array} \right)  \right .\\
&& \left .
\left ( \begin{array}{cc}  m_{ij}^{(1)}  \left [ 1 - \frac{E_+ (E_+-\epsilon_{y}) }{E_+^{(\Delta)}(E_+-E_-)} + \frac{E_- (E_--\epsilon_{y}) }{E_-^{(\Delta)}(E_+-E_-)} \right ] &
m_{ij}^{(12)} \frac{\epsilon_{xy}}{(E_+-E_-)} \left [ \frac{E_+  }{E_+^{(\Delta)}} - \frac{E_-  }{E_-^{(\Delta)}} \right ] \\
 m_{ij}^{(12)} \frac{\epsilon_{xy} }{(E_+-E_-)}\left [ \frac{E_+  }{E_+^{(\Delta)}} -
   \frac{E_-  }{E_-^{(\Delta)}} \right ] & m_{ij}^{(2)} \left [1 -  \frac{E_+ (E_+-\epsilon_{x}) }{E_+^{(\Delta)}(E_+-E_-)} + \frac{E_- (E_--\epsilon_{x}) }{E_-^{(\Delta)}(E_+-E_-)} \right ]
\end{array} \right)
\left ( \begin{array}{cc} \theta_1(-q)\\ \theta_2 (-q) \end{array} \right)  \right \} \nonumber
\ea
where $m_{ij}^{(\alpha)} \equiv \frac{\partial^2
  \epsilon_{\alpha}}{\partial k_i \partial k_j} $,  and we define $\varphi(q) = \theta_1(q) -\theta_2(q)$.  We have dropped the linear term in $\Omega$, because it is a total derivative and hence should not contribute to the action.  Here all $\epsilon, E,$ and $\Delta$ are evaluated at the momentum $k$ to be integrated over.  Note that we have also included the quadratic terms in the last line of
Eq. (\ref{Eq_BCSBasis}).

Evaluating $Tr (G_0 \Sigma G_0 \Sigma)$ gives:
\ba  \label{M2}
\frac{1}{2} Tr (G \Sigma G \Sigma) &=& - \frac{\Omega^2}{8} \left ( \begin{array}{cc} \phi(q) & \varphi(q) \\ \end{array} \right)
\left ( \begin{array}{cc} N_{\phi\phi} & 0 \\ 0 & N_{\varphi \varphi} \\
\end{array} \right)
\left ( \begin{array}{cc}\phi(-q) \\ \varphi(-q) \end{array} \right)  \n
&& +\int \frac{d^2 k}{ (2 \pi)^2} \left \{
  \frac{\epsilon_{xy}^2}{4} \varphi(q)
  \varphi(-q)\frac{-E_+^{(\Delta)}E_-^{(\Delta)}+\Delta^2+E_+E_-}{
    E_+^{(\Delta)}E_-^{(\Delta)} (E_+^{(\Delta)}+E_-^{(\Delta)})}
\right \}\n
&&+ \frac{1}{8}\frac{ 2 q_i q_j}{(E_+^{(\Delta)}+E_-^{(\Delta)} )(E_+-E_-)^2}\left ( 1-\frac{\Delta^2+E_+ E_-}{(E_+^{(\Delta)} E_-^{(\Delta)}} \right )
\left ( \begin{array}{cc} \phi(q) \varphi (q) \end{array} \right) \n
&& \left \{
\left ( \begin{array}{cc}
 -\left( \epsilon_{xy}^2+\frac{1}{2} ( \epsilon_{x} -
   \epsilon_{y} )^2 \right)  v^{(xy)}_i v^{(xy)}_j  -
 \epsilon_{xy}^2 v_{i \phi} v_{j \phi} &
 \frac{\epsilon_{xy}^2}{2} \left( v_{i \phi} v_{j \varphi} + v_{j
     \phi} v_{i \varphi} \right ) \\
\frac{\epsilon_{xy}^2}{2} \left( v_{i \phi} v_{j \varphi} + v_{j
     \phi} v_{i \varphi} \right ) &  \left( \epsilon_{xy}^2-\frac{1}{2} ( \epsilon_{x} -
   \epsilon_{y} )^2 \right)  v^{(xy)}_i v^{(xy)}_j  -
 \epsilon_{xy}^2 v_{i \varphi} v_{j \varphi}
\\
\end{array} \right)
\right . \n && \left . +
\left ( \begin{array}{cc}
0 & - \frac{\epsilon_{xy} (\epsilon_{x} - \epsilon_{y}) }{4}
\left(v^{(xy)}_i v^{(\phi)}_j +  v^{(xy)}_j v^{(\phi)}_i \right ) \\
- \frac{\epsilon_{xy} (\epsilon_{x} - \epsilon_{y}) }{4}
\left(v^{(xy)}_i v^{(\phi)}_j +  v^{(xy)}_j v^{(\phi)}_i \right )  &
- \frac{\epsilon_{xy} (\epsilon_{x} - \epsilon_{y}) }{2}
\left(v^{(xy)}_i v^{(\varphi)}_j +  v^{(xy)}_j v^{(\varphi)}_i \right ) \\
\end{array} \right) \right. \n
&& \left .
+
\left ( \begin{array}{cc}
0 & \tilde{\Lambda}_{\phi \varphi} \\
 \tilde{\Lambda}_{\phi \varphi}   &
 \tilde{\Lambda}_{\varphi \varphi} \\ \end{array} \right)
\right \}
\left ( \begin{array}{cc} \phi(-q) \\ \varphi (-q) \end{array} \right) \n
 \ea
where $v_{\phi i } \equiv \partial (\epsilon_{x}+\epsilon_{y})
/\partial k_i, v_{\varphi i } \equiv \partial (\epsilon_{x}-\epsilon_{y}) /\partial k_i$.
Here $ \tilde{\Lambda}_{\alpha \beta} $ are terms which come from
expanding traces involving $B$ in Eq. \ref{Eq_ABCs} to quadratic order in $q$.

Combining the mass terms from Eqs.  (\ref{M1}) and (\ref{M2}) gives the total mass term:
\be \label{Eq_Mtot}
M = \int \frac{d^2 k}{ (2 \pi)^2} \left \{ \frac{4 \Delta^2
    \epsilon_{xy}^2}{ E_+^{(\Delta)}E_-^{(\Delta)}
    (E_+^{(\Delta)}+E_-^{(\Delta)})} - 2 \epsilon_{xy} \right \}
\equiv \int \frac{d^2 k}{ (2 \pi)^2} \frac{4 \Delta^2
    \epsilon_{xy}^2}{ E_+^{(\Delta)}E_-^{(\Delta)}
    (E_+^{(\Delta)}+E_-^{(\Delta)})}
\ee
where the second equality holds because in practice $\epsilon_{xy}$
averages to $0$ over the Brillouin zone.

In simplified form, the momentum-dependent terms are:
\ba
\left( \begin{array}{cc}
N_{\phi \phi } c^2_{\phi \phi, ij}
& c^2_{\phi \varphi, ij}  \\
c^2_{\phi \varphi, ij} &
N_{\varphi \varphi} c^2_{\varphi \varphi, ij} \\
\end{array} \right )
&=& \frac{ 1}{8} \int \frac{d^2 k}{ (2 \pi)^2} \left
  \{ -\frac{1}{(E_+-E_-)} \left [
  \frac{E_+  }{E_+^{(\Delta)}} - \frac{E_-  }{E_-^{(\Delta)}} \right
]\right . \n
&&
\left( \begin{array}{cc}
m_{ij}^{(xx)}  \epsilon_{x }- 2  m_{ij}^{(xy)}
  \epsilon_{xy} + m_{ij}^{(yy)} \epsilon_{y} &
  m_{ij}^{(xx)} \epsilon_{x }- m_{ij}^{(yy)} \epsilon_{y}  \\
  m_{ij}^{(xx)} \epsilon_{x }- m_{ij}^{(yy)} \epsilon_{y}
&        m_{ij}^{(xx)}  \epsilon_{x }+2  m_{ij}^{(xy)}
  \epsilon_{xy} + m_{ij}^{(yy)} \epsilon_{y}  \\
\end{array} \right )
\n
+ &&\frac{ 2}{(E_+^{(\Delta)}+E_-^{(\Delta)} )(E_+-E_-)^2}\left ( 1-\frac{\Delta^2+E_+ E_-}{(E_+^{(\Delta)} E_-^{(\Delta)}} \right )\n
&& \left [
\left( \begin{array}{cc}
 -\left( \epsilon_{xy}^2+\frac{1}{2} ( \epsilon_{x} -
   \epsilon_{y} )^2 \right)  v^{(xy)}_i v^{(xy)}_j  -
 \epsilon_{xy}^2 v_{i \phi} v_{j \phi} &
 \frac{\epsilon_{xy}^2}{2} \left( v_{i \phi} v_{j \varphi} + v_{j
     \phi} v_{i \varphi} \right ) \\
\frac{\epsilon_{xy}^2}{2} \left( v_{i \phi} v_{j \varphi} + v_{j
     \phi} v_{i \varphi} \right ) &  \left( \epsilon_{xy}^2-\frac{1}{2} ( \epsilon_{x} -
   \epsilon_{y} )^2 \right)  v^{(xy)}_i v^{(xy)}_j  -
 \epsilon_{xy}^2 v_{i \varphi} v_{j \varphi}
\\
\end{array} \right )
\right . \n && \left . +
\left( \begin{array}{cc}
0 & - \frac{\epsilon_{xy} (\epsilon_{x} - \epsilon_{y}) }{4}
\left(v^{(xy)}_i v^{(\phi)}_j +  v^{(xy)}_j v^{(\phi)}_i \right ) \\
- \frac{\epsilon_{xy} (\epsilon_{x} - \epsilon_{y}) }{4}
\left(v^{(xy)}_i v^{(\phi)}_j +  v^{(xy)}_j v^{(\phi)}_i \right )  &
- \frac{\epsilon_{xy} (\epsilon_{x} - \epsilon_{y}) }{2}
\left(v^{(xy)}_i v^{(\varphi)}_j +  v^{(xy)}_j v^{(\varphi)}_i \right ) \\
\end{array} \right )  \right. \n
&& \left . \left.
+
\left( \begin{array}{cc}
0 & \tilde{\Lambda}_{\phi \varphi} \\
 \tilde{\Lambda}_{\phi \varphi}   &
 \tilde{\Lambda}_{\varphi \varphi} \\
\end{array} \right )
\right ] \right \}  \ \ \ .
\ea

\section{Effective action with Coulomb terms} \label{CoulombApp}

Including terms
generated by the Coulomb repulsion modifies the interaction term $\Sigma$
of the full fermion
propagator (\ref{Eq_ForSigma}) according to\cite{SharapovBeck} :
\be \label{Eq_SWithChi}
 \Sigma = \tau_3 \otimes \mathbf{1} ( i \frac{i\Omega \phi}{2} - i e \chi ) + \tau_3 \otimes \tau_3 i \frac{\Omega \varphi}{2}   + \Sigma'_{kin}
\ee
where $\Sigma_{kin}$ involves only spatial derivatives of the phases
$\theta_1$ and $\theta_2$.   Here $\chi$ is the Hubbard-Stratonovich field associated with the
Coulomb interaction.  The form of the coupling for $\chi$  to fermions
can be deduced from gauge invariance:
the phase $\theta_i$ is obviously a gauge-dependent quantity, and the
gauge-invariant degrees of freedom are the combinations $
\partial_\tau \theta/2 - e \chi - e A_0$ and $
\mathbf{\bigtriangledown} \theta /2 - e/c \mathbf{A}$
\cite{SharapovBeck}.  Hence the effective action for $\chi$ is the
same as that for $\partial_\tau \theta/(2e)$.  Eq. (\ref{Eq_SWithChi}) indicates that $\chi$ couples
in all cases like the time derivative of the {\em symmetric} component of the phase fluctuations.

To obtain the full effective action for the phase only modes in the
presence of Coulomb interactions, we first integrate out the
fermions, giving an effective action for the $3$
Hubbard-Stratonovich fields $\phi, \varphi, \chi$.
There are two relevant contributions: from $ Tr G \Sigma$, we obtain:
\be
- i e \chi(q) Tr \left [ (\tau_3 \otimes 1) G_{k-q} \right ] = - i e
\chi(q) \langle\rho_{k-q} \rangle
\ee
which cancels the first-order term in $\chi$ in the effective action (\ref{Eq_FProp}).

From $Tr G \Sigma G \Sigma$, we obtain contributions whose
coefficients are {\it the same} as the contributions from the time
derivatives of $\phi$.  In particular,
as the coefficients of the cross-terms in $q, \Omega$ from traces $G
\Sigma G \Sigma$ all vanish, the couplings
between $\chi$ and $\phi, \varphi$ depend only on $\Omega$. Hence the effective action for the
fields $\phi, \varphi, \chi$  has the form:
\ba \label{Stot}
S_{eff} &=& \int d \Omega d^2  q \left ( \begin{array}{ccc} \phi &  \varphi & \chi\\\end{array} \right)
\left ( \begin{array}{ccc}
N_{\phi \phi}\left[ \Omega^2- c_{\phi\phi}^2 q^2 \right ] & c_{\phi \varphi}^2 (q_x^2-q_y^2) & - 2 \Omega N_{\phi \phi} \\
c_{\phi \varphi}^2 (q_x^2-q_y^2) & N_{\varphi \varphi}\left[ \Omega^2
  - \Omega_0^2 - c_{\varphi \varphi}^2 q^2\right] & 0 \\
 -2 \Omega N_{\phi \phi} &0 & U^{-1}(q) + 4 N_{\phi \phi} \\
\end{array} \right)
\left ( \begin{array}{c} \phi \\  \varphi \\ \chi \\
 \end{array} \right)  \n
&&+ \Delta_\alpha V^{-1} _{\alpha, \beta} \Delta_\beta \ea where the
coefficients $N, c$ are given in Eqs (\ref{Eq_Ntot}), (\ref{M1}),
and  (\ref{M2}). The dispersion relation is given by finding the
values of $q, \Omega$ at which $\mathcal{M}$ is singular. Depending
on the values of the parameters, $\mathcal{M}$  may have one or two
modes which are finite as $q  \rightarrow 0$.  One of these is the
gapped Leggett mode; the other is a sound-like mode (the
Carlson-Goldman mode) which we find to be absent at $T=0$,
consistent with Ref. \onlinecite{SharapovBeck}.  The third mode is,
of course, the plasma mode, which does not appear in the low-energy
spectrum. To study only the phase modes, we may equivalently
integrate out $\chi$ and $\phi$ to obtain Eq. (\ref{SeffCoul} ).

\end{widetext}

\end{appendix}

\bibliography{iron3}

\end{document}